\documentclass[sigconf]{acmart}
\AtBeginDocument{%
  }

\setcopyright{acmlicensed}
\copyrightyear{2026}
\acmYear{2026}
\acmDOI{XXXXXXX.XXXXXXX}
\acmConference[CCS'26]{ACM Conference}{November 2026}{The Hague, The Netherlands}

\usepackage{amsmath}
\usepackage{amsthm}
\usepackage{enumitem}
\usepackage{booktabs}
\usepackage{siunitx}

\newtheorem{theorem}{Theorem}[section]
\newtheorem{lemma}{Lemma}[section]

\begin{document}

\title{Proving Circuit Functional Equivalence in Zero Knowledge}


\author{Sirui Shen}
\affiliation{%
  \institution{Centrum Wiskunde \& Informatica}
  \city{Amsterdam}
  \country{Netherlands}
}
\email{sirui.shen@cwi.nl}

\author{Zunchen Huang}
\affiliation{%
  \institution{Centrum Wiskunde \& Informatica}
  \city{Amsterdam}
  \country{Netherlands}
}
\email{zunchen.huang@cwi.nl}

\author{Chenglu Jin}
\affiliation{%
  \institution{Centrum Wiskunde \& Informatica}
  \city{Amsterdam}
  \country{Netherlands}
}
\email{chenglu.jin@cwi.nl}


  
\begin{abstract}
The modern integrated circuit (IC) ecosystem is increasingly reliant on third-party intellectual property (3PIP) integration, which introduces security risks, including hardware Trojans, security bugs/vulnerabilities. Addressing the resulting trust deadlock between IP vendors and system integrators without exposing proprietary designs requires novel privacy-preserving verification techniques. 
However, existing privacy-preserving hardware verification methods are all simulation-based and therefore fail to offer formal guarantees. In this paper, we propose ZK-CEC, the first privacy-preserving framework for hardware formal verification. By combining formal verification and zero-knowledge proof (ZKP), ZK-CEC establishes a foundation for formally verifying IP correctness and security without compromising the confidentiality of the designs.

We observe that existing zero-knowledge protocols for formal verification are designed to prove statements of public formulas. However, in a privacy-preserving verification context where the formula is secret, these protocols cannot prevent a malicious prover from forging the formula, thereby compromising the soundness of the verification.
To address these gaps, we first propose a general blueprint for proving the unsatisfiability of a secret design against a public constraint, which is widely applicable to proving properties in software, hardware, and cyber-physical systems. Based on the proposed blueprint, we construct ZK-CEC, which enables a prover to convince the verifier that a secret IP's functionality aligns perfectly with the public specification in zero knowledge, revealing the size of the proof and the gate-count of the IP. We implement ZK-CEC and evaluate its performance across various circuits, including arithmetic units and cryptographic components. Experimental results show that ZK-CEC successfully verifies practical designs, such as the AES S-Box, within practical time limits.
\end{abstract}

\ccsdesc[100]{Hardware~Equivalence checking}
\ccsdesc[500]{Security and privacy~Privacy-preserving protocols}


\keywords{Zero-knowledge Proofs, Hardware formal verification, Combinational equivalence check, Hardware supply chain security}


\maketitle
\section{Introduction}

With the rapid growth of IC technology, chips have become essential components in everything from personal equipment to large-scale infrastructure. To meet the evolving needs and manage increasing design complexity, the system-on-chip (SoC) paradigm has become the dominant business model~\cite{saleh2006system,kundu2018network}. In this model, system integrators (buyers) combine intellectual property (IP) cores into a single silicon die. This ecosystem builds a globalized supply chain consisting of IP Vendors and System Integrators.

Despite the efficiency, the increasing reliance on third-party IP (3PIP) introduces significant security risks to the hardware supply chain~\cite{rostami2014primer,ray2018protecting}. A malicious vendor may provide an IP core embedded with hardware Trojans~\cite{baumgarten2011case,ray2018protecting}, backdoors~\cite{waksman2011silencing}, or intentional logic bugs, leading to supply chain poisoning. Furthermore, malicious vendors might attempt to defraud buyers by providing forged designs. Conversely, if a buyer demands full transparency to verify the design, the vendor faces the risk of IP leakage before acquisition. Establishing a foundation of trust between vendors and buyers without compromising confidentiality has thus become a critical challenge in the hardware supply chain.

Existing research attempts to address the trust issue by privacy-preserving IP verification, which allows a prover to convince a verifier of an IP’s correctness without revealing the underlying design. One line of research uses homomorphic encryption (HE) to encrypt test vectors and outsource simulation to third parties~\cite{gouert2020romeo,konstantinou2015privacy}. Another research line leverages zero-knowledge proof (ZKP), where a verifier challenges a design encoded as a zero-knowledge virtual machine (zkVM) with test vectors to obtain verifiable simulation results~\cite{mouris2020pythia,mouris2022zk}.

However, the aforementioned approaches are all simulation-based, which at best increase confidence in the validated designs but cannot completely rule out the presence of malicious logic. In hardware verification, simulation alone is insufficient to capture corner cases, particularly stealthy hardware Trojans that are activated only under rare or carefully crafted conditions~\cite{jin2008hardware,rad2008power, haider2017advancing,haider2019advancing}. In contrast, formal verification provides a mathematical guarantee of correctness and security~\cite{love2011proof,jin2012proof}. Specifically, combinational equivalence checking (CEC)~\cite{goldberg2001using} ensures that a circuit is functionally equivalent to the golden model or specification by constructing a miter circuit and proving its unsatisfiability (UNSAT). In our framework, the functional specification is public and can be a truth table or an unoptimized reference design. While the functionality is public, the value of a highly optimized proprietary implementation still holds because the minimum circuit size problem is believed to be NP-hard~\cite{hitchcock2015np} (e.g., researchers are continually improving the implementations of the AES S-Box~\cite{reyhani2018smashing,maximov2019new}).

While recent studies~\cite{luo2022proving,luick2024zksmt,laufer2024zkpi,cuellar2023cheesecloth,kolesar2025coinductive} have proposed privacy-preserving formal methods for theorem proving and software analysis, their application to hardware is limited. Notably, ZKUNSAT represents the state-of-the-art in proving the unsatisfiability of a public formula in a zero-knowledge framework~\cite{luo2022proving}. However, we observe that ZKUNSAT cannot be directly applied to secret formula verification because it requires the formula statement to be public. Simply treating the formula as a hidden input would compromise the soundness of the protocol, allowing a malicious prover to prove properties about a forged formula.

To bridge this gap, we first propose a blueprint for privacy-preserving formal verification based on ZKUNSAT. Our approach involves partitioning a formula into the secret and the public parts. By checking constraints on the two parts, our blueprint ensures that the protocol achieves soundness against a malicious prover. This blueprint provides a generalized solution for various tasks requiring privacy-preserving formal verification.

Building on the blueprint, we present \textbf{the first zero-knowledge protocol for combinational equivalence checking}. By proving the functional equivalence between a secret IP and a public specification, our protocol guarantees functional correctness for the secret IP. We design four sub-protocols corresponding to the constraints in our blueprint. Based on these building blocks, we propose $\Pi_{\text{ZK-CEC}}$ and prove that it satisfies completeness, soundness, and zero-knowledge. Finally, we implement our protocol in the EMP-toolkit and evaluate its performance on circuits of different scales, and demonstrate its practical utility. We also provide an optimization with leakage analysis to improve the efficiency of our protocol, which speeds up the proving time by up to $2.88\times$.

\vspace{4pt}
\noindent
\textbf{Our contributions.} \par
\vspace{-4pt}
\begin{itemize}[leftmargin=*]
    \item We propose a generalized blueprint for privacy-preserving formal verification. By partitioning formulas into secret and public components and enforcing cross-constraints, we enable formal proofs over hidden logic while maintaining protocol soundness.
    \item Leveraging the proposed blueprint, we introduce the first zero-knowledge protocol for combinational equivalence checking, which allows vendors to prove that their secret IP is functionally equivalent to a public specification, providing a formal guarantee against logic-level Trojans and design errors.
    \item We implement our protocol and evaluate it under a set of common combinational circuits. Our evaluation provides a detailed analysis of the communication and computation overhead, with a detailed performance breakdown across different stages of the protocol to identify primary bottlenecks.
    \item We propose an optimization that merges specific steps within the proof to reduce the time overhead. We present a lower bound analysis showing that additional leakage introduced by the optimization does not enable formula reconstruction by the verifier. By implementing this optimization, we achieve up to a 2.88x speedup in proving time. 
\end{itemize}

\noindent
\textbf{Paper organization.}
The paper is organized as follows. Sec.~\ref{sec:set} establishes the problem setting and the threat model. Sec.~\ref{sec:pre} reviews the preliminaries, including Boolean logic resolution and VOLE-based zero-knowledge proofs. We propose our general blueprint for privacy-preserving property checking in Sec.~\ref{sec:blue}. Sec.~\ref{sec:prot} applies the blueprint to combinational equivalence checking, detailing the sub-protocols and the main protocol $\Pi_{ZK-CEC}$. Sec.~\ref{sec:impl} presents our implementation and evaluates the framework's performance across various circuit benchmarks. Finally, Sec.~\ref{sec:related} discusses related work, and Sec.~\ref{sec:conclusion} concludes the paper.
\section{Problem Setting and Threat Model}
\label{sec:set}


\noindent
\textbf{Problem setting.} In our problem setting, we consider a hardware IP vendor as the prover, and a prospective buyer as the verifier. Let $C_\text{spec}$ denote a public specification of a circuit. The prover holds a circuit implementation $C_\text{impl}$ and wants to convince the verifier that $C_\text{impl}$ is functionally correct or free of logic-level Trojan. The vendor commits $C_\text{impl}$ to the verifier and asserts its functional equivalence with $C_\text{spec}$. After the two parties run the zero-knowledge protocol, the verifier is convinced that $C_\text{impl} \equiv C_\text{spec}$. Finally, the prover opens the commitment, and the verifier pays the money via a fair exchange protocol.



\noindent
\textbf{Threat model.} We consider a malicious adversarial model where both the prover and the verifier may deviate from the protocol to gain an advantage. A malicious vendor may attempt to defraud the buyer by forging proof for an incorrect design. Conversely, a malicious buyer may attempt to learn information about the IP design before completing the trade.

\noindent
\textbf{Goal.} Our protocol aims to achieve completeness, soundness, and zero-knowledge. 
\begin{itemize}[leftmargin=*]
    \item \textbf{Completeness.} If the prover and the verifier run the protocol honestly on two equivalent designs $C_\text{impl}$ and $C_\text{spec}$, the two parties must accept.
    
    \item \textbf{Soundness.} 
    If $C_\text{impl}$ is not functionally equivalent to $C_\text{spec}$, the probability that a verifier accepts a proof from a malicious prover is $\text{negl}(\lambda)$, where $\lambda$ denotes the security parameter. This guarantees that no prover can forge a valid proof for an incorrect design, except with a negligible probability.
    
    \item \textbf{Zero-knowledge.} The verifier learns that $C_\text{impl} \equiv C_\text{spec}$ but nothing else about $C_\text{impl}$ except the publicly defined leakage below. 
\end{itemize}

\noindent
\textbf{Leakage.}
ZK-CEC discloses the quantity and maximum width of the used clauses, as well as the total number of the literals in the proof. The clause size encompasses both the number of resolution steps and the input clauses. Revealing the former is aligned with~\cite{luo2022proving}, which prevents the verifier from inferring information about the proof. The latter implies the gate-count of the circuit and is permissible to reveal in our scenario. It allows the prover to demonstrate the design's advantage in terms of gate-count, while the verifier still cannot reconstruct the circuit.

\section{Preliminaries}
\label{sec:pre}

\subsection{Boolean Logic and Resolution}
\label{subsec:reso}

In this work, we model hardware circuits using Boolean functions in propositional logic. Let $\mathcal{V} = \{x_1, x_2, \dots, x_n\}$ be a finite set of Boolean variables. We define the set of literals as $\mathcal{L} = \mathcal{V} \times \{0, 1\}$. Specifically, a literal $l = (x, b) \in \mathcal{L}$ is positive if $b=1$ and negative if $b=0$, conventionally denoted as $x$ and $\neg x$, respectively. 

A clause $c$ is a finite set of literals, logically representing the disjunction (OR) of these literals. For example, $c = \{x_1, \neg x_2\}$ corresponds to the logical expression $(x_1 \lor \neg x_2)$. A clause containing no literals is called the empty clause, denoted as $\bot$.

A Conjunctive Normal Form (CNF) formula $\Phi$ is defined as a finite collection of clauses, $\Phi = \{c_1, \dots, c_m\}$, representing the logical conjunction (AND) of these clauses $\Phi = c_1 \wedge \dots \wedge c_m$.

A Boolean assignment $\omega$ is a mapping that assigns a Boolean value to each variable, which defines an assigned literal set. An assignment $\omega$ satisfies a clause $c$ (denoted as $\omega \models c$) if and only if there exists at least one literal in $c$ that evaluates to 1 under $\omega$. 
 
The Boolean Satisfiability (SAT) problem asks whether there exists an assignment $\omega$ such that all clauses in $\Phi$ are satisfied simultaneously. If such an assignment exists, $\Phi$ is satisfiable; otherwise, $\Phi$ is unsatisfiable. In our work, the satisfiability of a formula is certified by providing a satisfying assignment $\omega$.

To certify unsatisfiability, we use the resolution proof system~\cite{robinson1965machine} to avoid enumerating of all possible assignments. 
Resolution is a sound and complete inference rule for propositional logic in CNF~\cite{davis1960computing}. Given two clauses $c_1 = (x \vee \alpha)$ and $c_2 = (\neg x \vee \beta)$, where $x$ is the pivot variable and $\alpha, \beta$ are disjunctions of literals, the resolution rule allows the derivation of a new clause called the resolvent:
$\text{Res}(c_1, c_2) = (\alpha \vee \beta)$. The resolvent is logically implied by its antecedents, i.e., any assignment $\omega$ that satisfies both $c_1$ and $c_2$ also satisfies the resolvent $(\alpha \vee \beta)$. Therefore, adding the resolvent to the formula preserves its satisfiability.

A refutation proof of UNSAT for a CNF formula $\Phi$ is a sequence of clauses $D_1, \dots, D_k$, where each $D_i$ is either a clause from the original formula $\Phi$ or a resolvent derived from two previous clauses in the sequence. The proof is successfully concluded when the sequence produces the empty clause $\bot$, which shows a logical contradiction, thereby proving that $\Phi$ is UNSAT. The length of a resolution proof corresponds to the total number of intermediate resolvents generated, which is equivalent to the number of resolution derivation steps performed to reach $\bot$.

\subsection{Combinational Equivalence Checking}

\begin{figure}[tbp]
\centerline{\includegraphics[width=5.6cm]{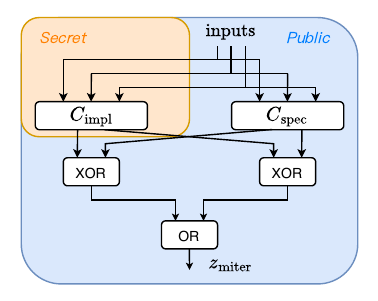}}
\caption{A miter circuit constructed by $C_\text{spec}$ and $C_\text{impl}$.}
\label{fig:miter}
\end{figure}

Combinational Equivalence Checking (CEC) involves determining whether two combinational circuits, typically a specification $C_\text{spec}$ and an implementation $C_\text{impl}$, have identical functional behavior. We formally treat these designs as Boolean functions mapping an input vector $X = (x_1, \dots, x_n)$ to an output vector $Y = (y_1, \dots, y_m)$, where $n$ and $m$ are the number of inputs and outputs, respectively. The two designs are considered logically equivalent if and only if for every possible input vector $X$, the output vectors are identical.

To verify this property, we can construct a miter circuit $C_\text{miter}$ as in Fig.~\ref{fig:miter} by tying the inputs of $C_\text{spec}$ and $C_\text{impl}$ together and comparing their corresponding outputs using XOR gates. All pairwise difference signals are aggregated using a multi-input OR gate to produce a single output $z_\text{miter}=\bigvee_{i=1}^{m}(y^{\text{spec}}_i \oplus y^{\text{impl}}_i)$. The signal $z_\text{miter}$ evaluates to $1$ if and only if the outputs of the two designs differ for some input. 
$C_\text{miter}$ contains three parts: $C_\text{spec}$, $C_\text{impl}$, and the output gates $C_\text{IO}$.

\begin{figure}[tbp]
\centerline{\includegraphics[width=6cm]{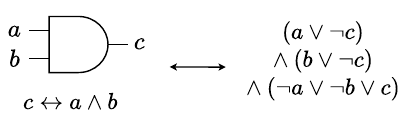}}
\caption{An AND gate and its corresponding CNF formula.}
\label{fig:AND-CNF}
\end{figure}

The Tseitin transformation~\cite{tseitin1983complexity} takes as input an arbitrary combinatorial circuit and produces an equi-satisfiable Boolean formula in CNF. Fig.~\ref{fig:AND-CNF} shows an example of a single AND gate and its corresponding CNF formula. Let $l_\text{out}$ be the literal representing the assignment $z_\text{miter} = \text{true}$. We translate $C_\text{miter}$ into a CNF formula and constrain the formula by asserting that $z_\text{miter}$ must be true: $\Phi_\text{miter}=\Phi_\text{spec}\land \Phi_\text{impl} \land \Phi_\text{IO}\land l_\text{out}$. With this approach, the problem of whether $C_\text{spec} \equiv C_\text{impl}$ can be decided by checking the unsatisfiability of $\Phi_\text{miter}$. If $\Phi_\text{miter}$ is unsatisfiable, it implies that no input can make the outputs differ. In this case, a resolution proof of unsatisfiability in Sec.~\ref {subsec:reso} serves as the formal certificate that $C_\text{spec}$ and $C_\text{impl}$ are equivalent.


\subsection{VOLE-based Zero-knowledge Protocols}

\begin{figure}[tbp]
    \centering
    \fbox{
        \begin{minipage}{0.95\linewidth}
            \vspace{0.5em}
            \centering
            \textbf{\underline{Functionality $\mathcal{F}_{\text{ZK}}$}}
            \vspace{0.5em}

            \raggedright 
                
                \textbf{Commitment:} On receiving $(\text{Commit}, x)$ from the prover, where $x \in \mathbb{F}$, store $x$ and send $[x]$ to each party.
                
                \textbf{Open:} On receiving $(\text{Open}, [x],y)$ from both parties, where $x,y \in \mathbb{F}$ and $[x]$ is a commitment of $x$. If the $x$ from the two parties does not match, or $x\neq y$, the functionality aborts.
                
                \textbf{Circuit relation:} On receiving $(\text{Relation}, C, [x_0], \dots, [x_{n-1}])$ from both parties, where $x_i \in \mathbb{F}$ and $C \in \mathbb{F}^n \rightarrow \mathbb{F}^m$, compute $y_1, \dots, y_m := C(x_0, \dots, x_{n-1})$ and send $\{[y_1], \dots, [y_m]\}$ to both parties.

                \textbf{Equal:} On receving $(\text{Equal}, [x_0], [x_1])$ from both parties, if $x_0 \neq x_1$ the functionality aborts.
        \end{minipage}
    }
    \caption{Functionality for ZKPs of elements in $\mathbb{F}$.}
    \label{fig:func_zk}
\end{figure}

Recent advancements~\cite{weng2021wolverine,yang2021quicksilver,baum2021mac,weng2021mystique,dittmer2020line} in zero-knowledge protocols rely on vector oblivious linear evaluation (VOLE) ~\cite{nielsen2012new} over a finite field $\mathbb{F}$. A VOLE is a two-party primitive involving a prover and a verifier. In this setup, the verifier holds a secret global key $\Delta \in \mathbb{F}$, chosen uniformly at random. For a vector of secrets $\mathbf{u} \in \mathbb{F}^n$ chosen by the prover, the VOLE protocol distributes randomness such that the prover obtains a vector $\mathbf{q} \in \mathbb{F}^n$ and the verifier obtains a vector of keys $\mathbf{v} \in \mathbb{F}^n$, satisfying the linear relation: $\mathbf{q} = \mathbf{v} + \mathbf{u} \cdot \Delta$.

The VOLE correlation can be viewed as an information-theoretic message authentication code (IT-MAC). For a single value $u$, the pair $(u, q)$ held by the prover serves as a commitment to $u$, authenticated by the verifier's key $(v, \Delta)$. The binding property ensures that the prover cannot open the commitment to a different value $u' \neq u$ without guessing $\Delta$. The verifier holds only the global key $\Delta$ and the local key $v$. As $v$ is uniformly distributed and independent of $u$, the verifier learns no information about the secret $u$. 

The VOLE-based commitment scheme supports efficient arithmetic operations due to the linearity of the correlation~\cite{sun2025committed}.
\begin{enumerate}[leftmargin=*]
\item \textbf{Linear Operations:} Given commitments to $x$ and $y$, denoted as $[x]$ and $[y]$, the parties can locally compute a commitment to any linear combination $z = \alpha x + \beta y + \gamma$ without interaction.
\item \textbf{Multiplication:} To verify a multiplicative relation $z = x \cdot y$, the protocols usually employ an interactive check. The prover and verifier interact to generate a new VOLE correlation for $z$ and subsequently prove $[z] = [x \cdot y]$.
\end{enumerate}

The VOLE-based zero-knowledge proofs follow the commit-and-prove paradigm. In this scheme, the prover first commits to the secret witness using the VOLE commitment. Then the prover proves to the verifier that the committed values satisfy the specific constraints of the computation. A functionality for zero-knowledge proofs of elements in $\mathbb{F}$ is shown in Fig.~\ref{fig:func_zk} to support arithmetization of CEC in zero-knowledge. Note that, in the paper, we treat the VOLE-based ZK backend as ideal functionalities, which is aligned with the existing work~\cite{luo2022proving,sun2025committed,luick2024zksmt}.

\subsection{Zero-knowledge Proof of Univariate Polynomials}

Our work arithmetizes the Boolean functions in univariate polynomials. In this paper, we present the ideal functionality of zero-knowledge proofs as $\mathcal{F}_\text{ZK-Poly}$ in Fig.~\ref{fig:func_zk_poly} based on $\mathcal{F}_\text{ZK}$. \textbf{Commit} is realized by committing every coefficient in $P(x)$ using $\mathcal{F}_\text{ZK}$. Evaluation on $chal$ is implemented using only linear operations of the VOLE commitment, which requires no interaction between the two parties.

In our protocol, we require verifying that the product of two sets of polynomials is identical. However, since $\mathcal{F}_\text{ZK-Poly}$ does not natively support polynomial multiplication, we cannot simply compute the product polynomial within the zero-knowledge domain and use \textbf{Equal} in $\mathcal{F}_\text{ZK}$. In $\mathcal{F}_\text{ZK-Poly}$, the parties can prove the identity of polynomials by applying the Schwartz-Zippel lemma. The lemma states that two non-zero distinct polynomials will evaluate to different values at a random point with overwhelming probability. To perform the check, the verifier generates a random challenge $chal$. The parties then evaluate the polynomials at $chal$. By computing the product of these evaluations for each set and comparing the results, they can verify with high probability that the product of the two sets of polynomials is identical.

For simplicity, in this paper, $[\Phi]$ and $[c]$ denote the commitments to all coefficients of the respective polynomial representations for formulas and clauses.

\begin{figure}[tbp]
    \centering
    \fbox{
        \begin{minipage}{0.95\linewidth}
            \vspace{0.5em}
            \centering
            \textbf{\underline{Functionality $\mathcal{F}_{\text{ZK-Poly}}$}}
            \vspace{0.5em}

            \raggedright 
                
                \textbf{Commit:} On receiving $(\text{Commit}, P(x))$ from the prover, where $P(x) \in \mathbb{F}[x]$, store $P(x)$ and send $[P(x)]$ to each party.
                
                \textbf{Open:} On receiving $(\text{Open}, [P(x)] , Q(x))$ from both parties, where $P(x),Q(x) \in \mathbb{F}[x]$ and $[P(x)]$ is a commitment of $P(x)$. If the $Q(x)$ from the two parties does not match, or $P(x)\neq Q(x)$, the functionality aborts.
                
                \textbf{Evaluation:} On receiving $(\text{Eval}, [P(x)], chal)$ from both parties, where $P(x) \in \mathbb{F}[x]$ and $chal \in \mathbb{F}$, if $chal$ from the two parties does not match, the functionality aborts; otherwise, compute $P(chal)\in \mathbb{F}$ and send $[P(chal)]$ to both parties.
                
                \textbf{Product-of-polynomials identity:} On receiving $(\text{PoPIdt}, \{[P_i(x)]\}_{i\in n}, \{[Q_j(x)]\}_{j\in m})$ from both parties, if $\prod_i\{[P_i(x)]\}\neq\prod_j\{[Q_j(x)]\}$, the functionality aborts.
                
        \end{minipage}
    }
    \caption{Functionality for ZKPs of univariate polynomials.}
    \label{fig:func_zk_poly}
\end{figure}

\subsection{ZKUNSAT}
\label{sec:zkunsat}
For the UNSAT proof component in our work, we adopt the protocol for checking resolution proof in~\cite{luo2022proving} (referred to as ZKUNSAT). In the ZKUNSAT framework, the protocol begins with the initialization of a public formula $\Phi$. The prover uses a solver to compute an UNSAT certificate of $\Phi$. The UNSAT certificate is structured as a sequence of resolution derivations, as described in Sec.~\ref{subsec:reso}. For each derivation in the sequence, the commitment to the involved clauses is operated through the functionality $\mathcal{F}_{clause}$ in Fig.~\ref{fig:func_clause}.

In ZKUNSAT, all clauses are encoded as univariate polynomials. Accordingly, the \textbf{Input} and \textbf{Equal} instructions in $\mathcal{F}_{clause}$ call the \textbf{Commit} and \textbf{PoPIdt} instructions of $\mathcal{F}_\text{ZK-Poly}$ to conduct clause commitment and equality verification. The \textbf{Res} instruction performs a single resolution step. Let $w_0$ and $w_1$ denote two witness clauses, and $p$ denote the pivot. A resolution $\{c_0, c_1\} \vdash c_r$ is transformed into the following constraints via clause weakening.
\begin{equation}
\label{eq:poly_relation}
\begin{split}
    w_0 \lor c_0 &= c_r \lor p \\
    w_1 \lor c_1 &= c_r \lor \neg p 
\end{split}
\end{equation}
For the pivots encoded as polynomials, the two parties must perform an additional check to ensure that the corresponding polynomial represents exactly one literal. Furthermore, the literal encoding scheme must constrain the relationship between a literal and its negation. By implementing these constraints, the resolution derivation in the polynomial encoding is proved to be both sound and complete~\cite{luo2022proving}, while overcoming the deduplication issue.

\begin{figure}[tbp]
    \centering
    \fbox{
        \begin{minipage}{0.95\linewidth}
            \vspace{0.5em}
            \centering
            \textbf{\underline{Functionality $\mathcal{F}_{\text{Clause}}$}}
            \vspace{0.5em}

            \raggedright 
                
                \textbf{Input:} On receiving $(\text{Input}, l_0, \cdots , l_{k-1}, w)$ from the prover and  $(\text{Input}, w)$ from the verifier where $l_i \in \mathcal{L}$, the functionality checks that $k \leq w$ and abort if it does not hold. Otherwise, store $c = l_0 \lor \cdots \lor l_{k-1}$, and send $[c]$ to each party.
                
                \textbf{Equal:} On receiving $(\text{Equal}, [c_0 ], [c_1])$ from the two parties, check if  $c_0 = c_1$; if not, the functionality aborts.
                
                \textbf{Res:}  On receiving $(\text{Res}, [c_0], [c_1], [c_r])$ from the two parties, check if $\{c_0, c_1\} \vdash c_r$ ; if not the functionality aborts.

                \textbf{IsFalse:} On receiving $(\text{IsFalse}, [c])$ from the two parties, check if $c=\bot$;  if not, the functionality aborts. 
        \end{minipage}
    }
    \caption{Functionality for clause operations in ZK.}
    \label{fig:func_clause}
\end{figure}

The sequence of resolution steps can be modeled as a RAM (random-access memory) program executing only one \textbf{Res} instruction. For further optimization, the prover can pre-store all intermediate clauses $[c_r]$ in memory. By validating the result of each derivation step against these pre-stored entries, the RAM program execution is reduced to a ROM (read-only memory) consistency check. The ROM in zero-knowledge is instantiated as the functionality $\mathcal{F}_\text{ZK-ROM}$ in Fig.~\ref{fig:func_zk_rom}. The prover and verifier first send all commitments of the input and intermediate clauses to initialize the ZK-ROM. During a resolution step, the two parties get fresh commitments of fetched clauses in the ZK-ROM and check if they comply with the resolution constraints. After the ROM being accessed by $t_{th}$ times, the two parties use \textbf{Check} instruction to check if those fetched commitments are consistent with the pre-stored commitments by the permutation check proposed in~\cite{franzese2021constant}.

\begin{figure}[tbp]
    \centering
    \fbox{
        \begin{minipage}{0.95\linewidth}
            \vspace{0.5em}
            \centering
            \textbf{\underline{Functionality $\mathcal{F}_{\text{ZK-ROM}}$}}
            \vspace{0.5em}

            \raggedright 
                
                \textbf{Initialization:} On receiving $(\text{Init}, N, [m_0 ], . . . , [m_{N-1} ])$ from the prover and the verifier, where $m_i \in \mathbb{F}$, store the $\{m_i\}$ and set $f :=  \text{honest}$  and ignore subsequent initialization calls.
                
                \textbf{Read:} On receiving $(\text{Read}, i, d, t )$ from the prover, and $(\text{Read}, t )$ from the verifier, where $d\in \mathbb{F}$ and $i,t \in \mathbb{N}$, send $[d]$ to each party. If $d\neq m_i$ or $t$ from both parties do not match or $i\geq t$, then set $f :=  \text{cheating}$.
                
                \textbf{Check:}  Upon receiving $(\text{Check})$ from the verifier, do: If the prover sends $(\text{cheat})$ then send $\text{cheating}$ to the verifier. If the prover sends $(\text{continue})$, then send $f$ to the verifier.                
                
        \end{minipage}
    }
    \caption{Functionality for read-only memory in ZK.}
    \label{fig:func_zk_rom}
\end{figure}

Finally, the two parties verify whether the last derived clause is the empty clause $\bot$ by calling \textbf{IsFalse} in $\mathcal{F}_\text{Clause}$. According to Theorem 4.7 in~\cite{luo2022proving}, the ZKUNSAT protocol is a zero-knowledge proof of knowledge of refutation proof. The Verifier is thus convinced of the unsatisfiability of the public formula (i.e., $\Phi$ is unsatisfiable) with only knowing the existence of a valid refutation proof.
\section{A Blueprint for Privacy-Preserving Property Checking}
\label{sec:blue}
\subsection{A Dilemma in Applying ZKUNSAT}

\begin{table*}[tbp]
  \centering
  \caption{Comparison between ZKUNSAT and Our Proposed Blueprint}
  \label{tab:comparison}
  \begin{tabular}{llll}
    \toprule
    \textbf{Feature} & \textbf{ZKUNSAT (Original)} & \textbf{ZKUNSAT (Modified)} & \textbf{Our Blueprint} \\
    \midrule
    \textbf{Formula $\Phi$} & Entirely Public & Entirely Secret & Private System + Public Property \\
    \textbf{Utility} & Low & Unknown & \textbf{High} \\
    \textbf{Soundness in property checking} & Yes & No & \textbf{Yes} \\
    \textbf{Use case} & Proof of co-NP statement & Secret program verification & Secret HW and SW verification \\
    \bottomrule
  \end{tabular}
\end{table*}

While ZKUNSAT suggests "proving the safety of secret programs" as a potential application, we identify two issues when deploying this protocol in practical property verification.

As the original ZKUNSAT protocol treats $\Phi$ as public, the primary utility of zero-knowledge vanishes. Since UNSAT belongs to co-NP, a verifier could simply execute a state-of-the-art SAT solver to check the formula directly. We do notice that under a specific problem construction in~\cite{luo2022proving}, checking refutation proof in ZK can be faster than SAT solving. However, in practical scenarios, the computational overhead of verifying a refutation proof in zero-knowledge does not show enough advantage in efficiency, leaving the protocol not well-motivated in public settings where secrecy is not required.

If the protocol is modified to treat $\Phi$ as a private input, which is consistent with the claimed application of secret program safety, it faces a critical soundness vulnerability. Because the verifier has no knowledge of the problem statement, it cannot ensure that $\Phi$ faithfully represents the semantics of the program being verified. For instance, a malicious prover could trivially append contradictory clauses (e.g., $(x) \land (\neg x)$) to a satisfiable formula. While the prover can still demonstrate "knowledge of a refutation proof", the proof is vacuous as it no longer correlates with the property of the original secret program.

Consequently, ZKUNSAT is trapped in a dilemma: in the original public formula setting, it is challenged by direct SAT solving; in a potential private formula setting, it loses its soundness due to the absence of a semantic correspondence check. To address this issue, our proposed protocol introduces a partially secret formula setting, which bridges the construction of ZKUNSAT and practical verification problems. Furthermore, the follow-up works~\cite{luick2024zksmt,karthikeyan2025towards} in this research line did not explicitly discuss the secrecy of the formula, and the dilemma still exists.

\subsection{A Blueprint for Secret Formula Property Checking}
\label{sec:blutprint}

\begin{figure}[tbp]
\centerline{\includegraphics[width=6.5cm]{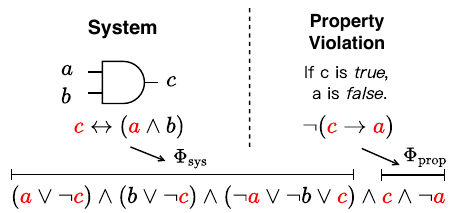}}
\caption{An AND-gate example of secret formula property checking. }
\label{fig:cnf}
\end{figure}

To resolve the aforementioned dilemma, we propose a refined setting for secret formula property checking. In this framework, the formula $\Phi_\text{sys}$ representing the prover's system remains secret, while the property formula $\Phi_\text{prop}$ to be verified is public. The verification goal is to demonstrate that the conjunction of the private system model and the public property violation condition is unsatisfiable. Fig.~\ref{fig:cnf} gives a toy example of verifying a property of an AND gate, where the variables colored in red are shared by both formulas.

We formally define the secret formula property checking problem. Let $\Phi_\text{sys}$ be a secret CNF formula that represents the system in question, and $\Phi_\text{prop}$ be a public property CNF formula. The formulas are defined over the variable sets $\mathcal{V}_\text{sys}$ and $\mathcal{V}_\text{prop}$, respectively. The goal is to provide a zero-knowledge proof that the conjunction of the secret system and the public property is unsatisfiable.
\begin{equation}
\Phi_\text{sys} \land \Phi_\text{prop} \models \bot
\end{equation}

We also define the variable set $\mathcal{V}_\text{IO} = \mathcal{V}_\text{sys} \cap \mathcal{V}_\text{prop}$, representing the shared interface between the system and the property. The rest part in $\mathcal{V}_\text{sys}$ is defined as $\mathcal{V}_\text{sec}$, where $\mathcal{V}_\text{IO} \cup \mathcal{V}_\text{sec} = \mathcal{V}_\text{sys}$.

To ensure the soundness of the protocol against a malicious prover, the verifier must check the following properties:
\begin{enumerate}[leftmargin=*]
    \item \textbf{Property correctness:} $\Phi_\text{prop}$ correctly describes the property, which also implies $\Phi_\text{prop}$ is satisfiable.
    \item \textbf{System satisfiability:} $\Phi_\text{sys}$ is satisfiable.
    \item \textbf{Variable isolation} $\mathcal{V}_\text{sec} \cap \mathcal{V}_\text{prop} = \emptyset$.
\end{enumerate}
The property correctness check ensures the validity of the statement, verifying that it corresponds to the security or functional requirements. The system satisfiability check ensures that the prover's secret does not contain any contradiction. The variable isolation ensures that $\Phi_\text{sys}$ and $\Phi_\text{prop}$ only contradict through the explicitly defined interface $\mathcal{V}_\text{IO}$. Lemma~\ref{lemma:sat} states that when the conjunction of two satisfiable formulas is unsatisfiable, the projections of their satisfying assignment sets on the shared variables are disjoint. A proof of Lemma~\ref{lemma:sat} is available in Appendix~\ref{sec:prf_lemma_sat}.

\begin{lemma}
\label{lemma:sat}
Let $\phi_a$ and $\phi_b$ be two satisfiable formulas with variable sets $\mathcal{V}_a$ and $\mathcal{V}_b$, respectively. We define $\mathcal{V}'_a=\mathcal{V}_a\setminus \mathcal{V}_b$ and $\mathcal{V}'_b=\mathcal{V}_b\setminus \mathcal{V}_a$. If $\phi_a \wedge \phi_b$ is unsatisfiable, then the sets of assignments on $\mathcal{V}_a \cap \mathcal{V}_b$ are disjoint:
\begin{equation*}
    \{ \omega \mid \omega \models \exists \mathcal{V}'_a\phi_a \} \cap \{ \omega \mid \omega \models \exists \mathcal{V}'_b\phi_b \} = \emptyset
\end{equation*}
\end{lemma}

As established by Lemma~\ref{lemma:sat}, properties (2) and (3) ensure that the contradiction proved in our protocol is rooted in the conflict between the two satisfiable solution spaces. Specifically, $\Phi_\text{sys}$ and $\Phi_\text{prop}$ project constraints onto $\mathcal{V}_{IO}$, defining two distinct sets of valid partial assignments $\omega_\text{sys}$ and $\omega_\text{prop}$. The unsatisfiability of the conjunction $(\Phi_\text{sys} \land \Phi_\text{prop})$ is a consequence of $\omega_\text{sys}$ and $\omega_\text{prop}$ being disjoint. In other words, the refutation proof serves as a formal certificate that no behavior of the secret system exposed via $\mathcal{V}_\text{IO}$ can satisfy the requirements of the public property. 

By conducting the property checks in zero-knowledge and composing them with ZKUNSAT, we establish a blueprint for privacy-preserving secret formula property checking. Our blueprint transforms ZKUNSAT from a theoretical primitive into a practical tool. Table~\ref{tab:comparison} summarizes the differences between the ZKUNSAT and our proposed blueprint regarding their utility. Our blueprint allows a prover to convince a verifier that its black-box system satisfies a public property without revealing the underlying implementation details. Our proposed blueprint provides a framework for privacy-preserving verification, adaptable to hardware, software, or general model checking. To demonstrate its practical utility, we apply our framework to combinational equivalence checking in our paper.

\section{Arithmetization and Protocol}
\label{sec:prot}

\subsection{Overview of Our Protocol}
ZK-CEC achieves correctness, soundness, and zero-knowledge, meaning both the prover and the verifier may be malicious. The goal of ZK-CEC is to check the equivalence of two circuits $C_\text{spec}$ and $C_\text{impl}$ by proving UNSAT of the formula $\Phi_\text{miter}$ representing the miter circuit. However, as mentioned in Sec.~\ref{sec:zkunsat}, simply applying ZKUNSAT~\cite{luo2022proving} for proving UNSAT of a secret formula lacks soundness. Hence, we achieve soundness in ZK-CEC by utilizing the structure of the CEC problem itself. 

The formula $\Phi_\text{miter}=\Phi_\text{spec}\land \Phi_\text{impl} \land \Phi_\text{IO}\land l_\text{out}$ can be divided into a public formula $\Phi_\text{pub}=\Phi_\text{spec} \land \Phi_\text{IO}\land l_\text{out}$ and a secret formula $\Phi_\text{sec}=\Phi_\text{impl}$. Aligning with the proposed blueprint in Sec.~\ref{sec:blutprint}, we ensure the correctness and soundness of proving UNSAT of $\Phi_\text{miter}$ by checking the following properties:
\begin{enumerate}[leftmargin=*]
    \item \textbf{P1. Correctness and Satisfiability of $\Phi_\text{pub}$.} The public clauses must follow the specification and the structure of a miter circuit, and they must be satisfiable.
    \item \textbf{P2. Unsatisfiability of $\Phi_\text{miter}$.} The miter circuit must be unsatisfiable.
    \item \textbf{P3. Satisfiability of $\Phi_\text{sec}$.} The formula corresponding to the secret circuit must be satisfiable, which means it cannot derive any local contraction.
    \item \textbf{P4. Non-intersection of $\Phi_\text{pub}$ and $\Phi_\text{sec}$.} The literals and their negations in the public clauses and private clauses must not have any intersection except the input and output.
\end{enumerate}
P1 is ensured by opening the commitments, which is discussed in Sec.\ref{sec:prove_corr}. The other properties are checked under arithmetization under the encoding scheme described in Sec.~\ref{sec:enc} that processes the public and the secret literals separately. For P2, Sec.~\ref{sec:prove_unsat} adopts the solution in~\cite{luo2022proving} that checks the UNSAT as a RAM program. Sec.~\ref{sec:prove_sat} discusses our approach to proving P3 using the same commitments in P2. We present another building block to prove P4 in Sec.~\ref{sec:non_inter} to ensure the separation of $[\Phi_\text{pub}]$ and $[\Phi_\text{sec}]$.  Finally, Sec.~\ref{sec:m_protocol} puts everything together and proposes the complete protocol for CEC in zero-knowledge.

In our framework, the arithmetization of elements and operations is performed over the finite field $\mathbb{F}_{2^k}$. We refer to this field simply as $\mathbb{F}$ in the remainder of this paper.

\subsection{Encoding of Formula}
\label{sec:enc}
To put all the proof checking into a zero-knowledge protocol, we need to transform the logic deduction into an arithmetic system. A CNF formula can be represented as a set of clauses, and a clause can be represented as a set of literals. The propositional calculus can be mapped into set operations. 
In our protocol, we encode clauses into univariate polynomials as in~\cite{perrucci2007real, luo2022proving}, where the literals are represented as different roots of polynomials. In this method, we convert the SAT problem to the existence of specific roots, and the set operations can be checked by polynomial operations. The encoding scheme in our framework adopts the approach in~\cite{luo2022proving}.

Let $c$ denote a clause of $n$ literals in a CNF formula. $\mathcal{L}$ denotes the set of all literals. $l_i\in \mathcal{L}$ denotes the $i$-th literal in the clause $c$. We define an encoding $\text{Poly}_{\phi}:\text{clause} \to \mathbb{F}[x]$:
\begin{equation*}
    \text{Poly}_{\phi}[l_0 \lor l_1 \lor \cdots \lor l_n] = (x-\phi(l_0))(x-\phi(l_1))\cdots (x-\phi(l_n))
\end{equation*}
that represents a clause as a univariate polynomial over $\mathbb{F}$. Each literal is encoded to an element in $\mathbb{F}$ under $\phi$ and embedded as a root of the polynomial. We give an example of converting a circuit and a property into encoded polynomials and committing those polynomials in Appendix~\ref{sec:app_example}. For brevity, we denote the polynomial corresponding to the encoding of clause $c$ as $\text{Poly}[c]$ in the remainder of this paper.

Through this encoding, we transform a set of literals into a set of roots. Consequently, the following properties can be verified via polynomial operations:
\begin{enumerate}[leftmargin=*]
    \item \textbf{Membership test:} A literal belongs to a clause if and only if its corresponding value is a root of the encoding polynomial.
    \begin{equation}
    \label{eq:member}
        l\in c \iff \text{Poly}[c](\phi(l))=0
    \end{equation}
    \item \textbf{Entailment check:} A clause $c$ is a sub-clause of the clause $c'$ if and only if the polynomial corresponding to $c$ divides the polynomial corresponding to $c'$.
    \begin{equation}
        c\to c' \iff \text{Poly}[c]\big| \text{Poly}[c']
    \end{equation}
    \item \textbf{Unique disjunction:} The conjunction of two disjoint clauses corresponds to the product of their encoded polynomials. 
    \begin{equation}
        \forall c_1,c_2,\ c_1\land c_2 =\emptyset \implies  \text{Poly}[c_1] \cdot \text{Poly}[c_2] = \text{Poly}[c_1 \lor c_2]
    \end{equation}
\end{enumerate}
Note that, in this polynomial encoding, deduplication of the union of two clauses is represented as the removal of the repeated roots, which introduces huge computational overhead in the ZKP backend. Therefore, we only check the disjunction of two disjoint clauses in our protocol.

Following the clause encoding, we define the literal encoding $\phi: \text{literal}\to \mathbb{F}$. Let $w_\text{lit}$ denote the maximum bitwidth of an encoded literal. We define two constants $const_\text{pub}=(1 \ll (w_\text{lit}-1))_\mathbb{F}$ and $const_\text{lit}=(1 \ll (w_\text{lit}-2))_\mathbb{F}$ that point to the two most significant bits (MSBs) of an encoded literal.

First, we establish the encoding constraint regarding a literal and its negation:
\begin{equation}
\label{eq:constraint}
    \phi(x) + \phi(\neg x) = const_\text{lit}
\end{equation}
This constraint enables us to check the relationship between complementary literals using addition. Additionally, under Eq.~\ref{eq:constraint}, we need to define only the positive literals, since their negations are naturally determined.

Based on the constraint of Eq.~\ref{eq:constraint}, we design a new adaptive literal encoding that maps secret and public literals to field elements separately.
In our protocol, the formula consists of $\Phi_\text{pub}$ and $\Phi_\text{sec}$. 
Let $\mathcal{L}_{\text{pub}}$ denote the set of literals corresponding to the variables in $\Phi_\text{pub}$, and let $\mathcal{L}_{\text{impl}}$ denote the set of literals corresponding to the variables in $\Phi_\text{impl}$. We define the set of secret literals $\mathcal{L}_{\text{sec}}=\mathcal{L}_{\text{impl}}\setminus \mathcal{L}_{\text{pub}}$.
To map literals into field elements, we define the mapping $\text{idx}(\cdot): \mathcal{L}_\text{pos} \to \{1, \dots, n\}$ that assigns a unique index to every positive literal, where $n$ is the number of positive literals.
Let $H_k(\cdot)$ be a collision-resistant keyed hash function with key $k$. The output bit-length of $H_k(\cdot)$ is $w_\text{lit}-1$ and smaller than that of $\mathbb{F}$; therefore, we can directly map the output bitstring into $\mathbb{F}$. We define $\phi$ as a deterministic mapping from positive literals to $\mathbb{F}$ as follows:
\begin{itemize}
    \item For $l \in \mathcal{L}_{\text{pub}}$, $\phi(l) = \text{idx}(l)+const_\text{pub}$.
    \item For $l \in \mathcal{L}_{\text{sec}}$, $\phi(l) = H_k(\text{idx}(l))$.
\end{itemize}
where the field elements of negative literals are defined by Eq.~\ref{eq:constraint}.

For public literals, we directly use the index as the encoded element, enabling both parties to compute it locally. We require the maximum index to be strictly less than $2^{w_{\text{lit}}-2}$, which ensures that the bit-flip introduced by $const_\text{lit}$ in Eq.~\ref{eq:constraint} does not cause a collision between the encoding of the negation $\phi(\neg l)$ and another positive literal $\phi(l')$. We set the MSB to 1 to mark the encoded literal as public, ensuring that public and secret encoded literals cannot collide.

For secret literals, the verifier should be prevented from determining their polarity by observing the encoded values. To achieve this, we use a collision-resistant keyed hash function $H_k(\cdot)$ for the encoding. 
The $H_k(\cdot)$ we use can be modeled as a pseudorandom function (PRF). Therefore, this scheme ensures that the polarity of a literal is indistinguishable from a random bit to the verifier, due to the indistinguishability of PRF. 
Furthermore, the collision resistance of $H_k(\cdot)$ ensures it is computationally infeasible for a malicious prover to find a key where two secret literals have collided encoding, i.e., $\phi(l) = \phi(l')$. The verifier's recomputation of the hash during commitment opening forces the prover to use the specified hash function.

\begin{lemma}
For any secret variable $x\in \mathcal{V}_\text{sec}$, and any adversary $\mathcal{A}$, the probability of distinguishing between $\phi(x)$ and $\phi(\neg x)$ is $\text{negl}(\lambda)$.
\end{lemma}

To prevent the leakage of clause width information in the proof, we ensure that all polynomials have the same length. We identify the maximum degree $w$ among polynomials and pad the high-order coefficients of shorter polynomials with zeros, which ensures that all polynomial commitments do not leak individual clause widths.

\subsection{Proving P1}

\begin{figure}[tbp]
    \centering
    \fbox{
        \begin{minipage}{0.95\linewidth}
            \vspace{0.5em}
            \centering
            \textbf{\underline{Protocol $\Pi_\text{P1}$}}
            \vspace{0.5em}

            \raggedright 
                \textbf{Parameters:} A finite field $\mathbb{F}$. A clause encoding scheme $\text{Poly}_\phi$ and a literal encoding scheme $\phi$.
                
                \textbf{Input:} Both parties have a public specification $C_\text{spec}$ and hold a commitment $[\Phi_\text{pub}]$ of all public clauses to check.
                
                \textbf{Protocol:}
                \begin{enumerate}[leftmargin=*]
                    \item The verifier constructs the IO $C_\text{IO}$ of the miter upon $C_\text{spec}$.
                    \item The verifier transform $C_\text{spec}$ and $C_\text{IO}$ to $\Phi'_\text{pub}$ using Tseytin transformation.
                    \item The prover sends the index assignment $\text{idx}(\cdot)$ of all literals in  $\Phi_\text{pub}$ to the verifier. The two parties compute $\mathcal{L}_\text{pub}$.
                    \item For $i$-th clause $c_i$ in  $\Phi'_\text{pub}$, the verifier checks the correctness by the following:
                    \begin{enumerate}
                        \item The verifier computes $P(x)=\text{Poly}_\phi[c_i]$ using $\text{idx}(\cdot)$.
                        \item Both parties send $(\text{Open}, [c_i],P(x))$ to $\mathcal{F}_\text{ZK-Poly}$. If the functionality aborts, the protocol aborts.
                    \end{enumerate}
                    
                \end{enumerate}
                
        \end{minipage}
    }
    \caption{Protocol for proving P1.}
    \label{fig:prot_p1}
\end{figure}

\label{sec:prove_corr}
In Fig.~\ref{fig:prot_p1}, we propose the protocol for the verifier to check the commitments of $\Phi_\text{pub}$. In this protocol, the prover first sends the index assignments for each literal within the arithmetized formula to the verifier. This enables the verifier to locally reconstruct the encoded polynomial. By opening the commitments, the verifier validates the correctness of the components corresponding to the public specification and the miter. Additionally, as a valid specification and miter, $\Phi_\text{pub}$ is naturally satisfiable.

\noindent
\textbf{Completeness.} The completeness of the protocol $\Pi_\text{P1}$ is guaranteed by the injectivity of $\text{Poly}_\phi$.

\noindent
\textbf{Soundness.} The soundness of the protocol $\Pi_\text{P1}$ is guaranteed by the binding property verified during the opening of the commitments.

\subsection{Proving P2}

\begin{figure}[tbp]
    \centering
    \fbox{
        \begin{minipage}{0.95\linewidth}
            \vspace{0.5em}
            \centering
            \textbf{\underline{Protocol $\Pi_\text{P2}$}}
            \vspace{0.5em}

                \raggedright 
                \textbf{Input:} The two parties have a committed formula $[\Phi]=[c_0] \land \cdots\land [c_{|\Phi|-1}]$ corresponding to the prover's secret. The prover has a key $k$ for the hash function $H$ in the literal encoding scheme $\phi$ and a refutation proof $((k_{l_0},k_{r_0},c_{|\Phi|}),\cdots,(k_{l_{R-1}},k_{r_{R-1}},c_{|\Phi|+R-1}))$; both parties know the length of the refutation proof $R$ and the maximum width of all clauses $w=\text{max}(c_i)$.
                
                \textbf{Protocol:}
                \begin{enumerate}[leftmargin=*]
                    \item The prover encodes $\{c_i\}_{i\in [|\Phi|,R+|\Phi|-1]}$ with $\text{Poly}_\phi(\cdot)$. The two parties send the Input instruction to $\mathcal{F}_\text{Clause}$ and obtain $[c_i]_{i\in [|\Phi|,R+|\Phi|-1]}$. 
                    \item The two parties send $(\text{Init}, R+|\Phi|-1, [c_0], \cdots , [c_{|\Phi|+R-1}])$ to $\mathcal{F}_\text{ZK-ROM}$.
                    \item For the $i$-th iteration, the two parties check the refutation proof by doing the following.
                    \begin{enumerate}
                        \item The prover looks up the tuple $(k_{l_i},k_{r_i},c_{|\Phi|+i})$ from the refutation proof such that $(c_{k_{l_i}},c_{k_{r_i}}) \vdash c_{|\Phi|+i}$.
                        \item The prover sends $(\text{Read}, k_{l_i}, c_{k_{l_i}}, i)$ to $\mathcal{F}_\text{ZK-ROM}$; the verifier sends $(\text{Read}, i)$ to $\mathcal{F}_\text{ZK-ROM}$, from which the two parties obtain $[c_{k_{l_i}}]$. Similarly, the two parties obtain $[c_{k_{r_i}}]$ and $[c_{|\Phi|+i}]$.
                        \item The two parties send $(\text{Res}, [c_{k_{l_i}}], [c_{k_{r_i}}], [c_{|\Phi|+i}])$ to $\mathcal{F}_\text{Clause}$.
                    \end{enumerate}
                    \item After $R$ iterations, two parties use $\mathcal{F}_\text{Clause}$ to check that $[c_{|\Phi|+R-1}]$ equals $\bot$; if $\mathcal{F}_\text{Clause}$ aborts, the verifier aborts.
                    \item Two parties send $(\text{Check})$ to $\mathcal{F}_\text{ZK-ROM}$, if $\mathcal{F}_\text{ZK-ROM}$ aborts, the verifier aborts.
                \end{enumerate}
                
        \end{minipage}
    }
    \caption{Protocol for proving P2.}
    \label{fig:prot_p2}
\end{figure}

\label{sec:prove_unsat}
To verify P2, we construct the protocol $\Pi_\text{P2}$ based on the main protocol introduced in~\cite{luo2022proving}. Unlike the original ZKUNSAT, $\Pi_\text{P2}$ operates on a committed formula, which replaces the initial commitment phase. Thus $\Pi_\text{P2}$ proves the unsatisfiability of a secret formula. $\Pi_\text{P2}$ assumes the refutation proof length $R$ and the maximum clause width $w$ are public. The detailed construction of $\Pi_\text{P2}$ is illustrated in Fig.~\ref{fig:prot_p2}.

In $\Pi_\text{P2}$, the prover has a refutation proof in the form of a tuple. There are three items in each element of the tuple, where $k_{l_i}$ and $k_{r_i}$ denote the indices of two antecedents in a resolution step, and $c_{|\Phi|+i}$ denotes the resolvent. All the polynomials corresponding to the resolvents are padded to the same length, hence their commitments $[c_{|\Phi|+i}]$ are also the same in length.

$\Pi_\text{P2}$ has three steps: First, the prover commits the resolvents, and the two parties initialize $\mathcal{F}_\text{ZK-ROM}$ with the commitments. Next, the two parties fetch clauses from $\mathcal{F}_\text{ZK-ROM}$ and verify the resolution steps. Each resolution step has two antecedents and one resolvent, and those clauses are hidden from the verifier as VOLE-based commitments. Note that all these clauses are fresh commitments fetched from the ZK-ROM; thus, the verifier remains unknown to their contents, regardless of whether the original clauses are public or secret. Finally, the two parties check that the last derived clause is empty and execute the permutation check of $\mathcal{F}_\text{ZK-ROM}$.

\noindent
\textbf{Completeness.} The completeness of the protocol $\Pi_\text{P2}$ holds because the protocol is deterministic.

\noindent
\textbf{Conditional soundness.} Note that $\Pi_\text{P2}$ does not guarantee the integrity of the input formula. Under the condition that the input formula is valid, the soundness of $\Pi_\text{P2}$ is reducible to the soundness of the main protocol in ZKUNSAT.


\subsection{Proving P3}

\begin{figure}[tbp]
    \centering
    \fbox{
        \begin{minipage}{0.95\linewidth}
            \vspace{0.5em}
            \centering
            \textbf{\underline{Protocol $\Pi_\text{P3}$}}
            \vspace{0.5em}

            \raggedright 
                \textbf{Parameters:} A finite field $\mathbb{F}$. A clause encoding scheme $\text{Poly}_\phi$ and a literal encoding scheme $\phi$.
                
                \textbf{Input:} Both parties hold a commitment $[\Phi_\text{sec}]$ of all secret clauses to be checked. The number of clauses in $[\Phi_\text{sec}]$ is $n$. The prover holds a valid assignment $\omega\models \Phi_\text{sec}$. Let $\mathcal{L}$ denote the set of literals corresponding to $\omega$. The prover also knows the number of literals $m$ in $\mathcal{L}$.
                
                \textbf{Protocol:}
                \begin{enumerate}[leftmargin=*]
                    \item The prover sends $m$ to the verifier.
                    \item For every $i \in \{0,1,\cdots,m-1\}$, the prover sends $\phi(l_i)$ to the verifier. 
                    \item The verifier checks for the presence of complementary literals. For every $i,j \in \{0,1,\cdots,m-1\}$, if $\phi(l_i) + \phi(l_j) = const_\text{lit}$, the protocol aborts.
                    \item The prover sends $(\text{Commit},0)$ to $\mathcal{F}_\text{ZK}$. The two parties receive $[0]$ and send $(\text{Open},[0], 0)$ to $\mathcal{F}_\text{ZK}$. If the functionality aborts, the protocol aborts.
                    \item For every $j \in \{0,1,\cdots,n-1\}$, the verifier checks the satisfiablity of $c_j\in \Phi_\text{sec}$ by the following:
                    \begin{enumerate}
                        \item For every $i \in \{0,1,\cdots,m-1\}$, both parties send $(\text{Eval},[c_j],l_i)$ to $\mathcal{F}_\text{ZK-Poly}$ and receive $[\text{Poly}[c_j](l_i)]$. 
                        \item Both parties send $(\text{PoPIdt},\{[\text{Poly}[c_j](l_i)]\}_{i\in m},\{[0]\})$  to $\mathcal{F}_\text{ZK-Poly}$. If the functionality aborts, the protocol aborts. \end{enumerate}
                    
                \end{enumerate}
                
        \end{minipage}
    }
    \caption{Protocol for proving P3.}
    \label{fig:prot_p3}
\end{figure}

\label{sec:prove_sat}
To check P3, the prover demonstrates to the verifier that $\Phi_\text{sec}$ is satisfiable without revealing its content. We propose a privacy-preserving protocol to prove SAT over our encoding scheme.

Let $\Phi_\text{sec}$ be a Boolean formula defined over a set of $n$ variables $\mathcal{V}_\text{impl} = \{x_1, x_2, \dots, x_n\}$. $\mathcal{V}_\text{impl}$ has two parts: $\mathcal{V}_\text{io}$ and $\mathcal{V}_\text{sec}$. The variables in $\mathcal{V}_\text{io}$ and $\mathcal{V}_\text{sec}$ correspond to the input/output wires and the internal wires of $C_\text{impl}$, respectively. Since $C_\text{impl}$ shares inputs with $C_\text{spec}$ and connects its outputs to $C_\text{IO}$, $\mathcal{V}_\text{IO}$ is known by both parties, implying $\mathcal{L}_\text{IO} \in \mathcal{L}_\text{pub}$. The internal wires of $C_\text{impl}$, represented by $\mathcal{V}_\text{sec}$, corresponds to $\mathcal{L}_\text{sec}$. Let $\omega_\text{sec}: \mathcal{V}_\text{sec} \to \{0, 1\}^n$ be a Boolean assignment. The SAT problem of $\Phi_\text{sec}$ is defined:
\begin{equation}
    \exists \omega_\text{sec},\ \forall c\in \Phi_\text{sec},\ \omega_\text{sec} \models c
\end{equation}

We transform the problem that $\omega_\text{sec}$ satisfies a clause $c$ into a membership testing problem as in Eq.~\ref{eq:member}; therefore, the verifier can verify P3 using polynomial arithmetic. We propose the protocol for proving P3 in Fig.~\ref{fig:prot_p3}.

\noindent
\textbf{Completeness.} 
When the two parties are honest, the completeness of the protocol $\Pi_\text{P3}$ is guaranteed by the completeness of the membership test in Eq.~\ref{eq:member} and the ZK backend. Provided that the assignment is valid, $\text{Poly}[c]$ vanishes at at least one encoded literal index $\phi(l_i)$, which implies that the product of its evaluations over the domain is zero.

\noindent
\textbf{Conditional soundness.} 
The soundness of the protocol $\Pi_\text{P3}$ is defined as: Assume that $\forall l\in \mathcal{L},\phi(l)+\phi(\neg l)=const_\text{lit}$, if $\Phi_\text{sec}$ is unsatisfiable, the verifier will accept that $\Phi_{\text{sec}}$ is unsatisfiable with negligible probability $\text{negl}(\lambda)$. A proof sketch is in Appendix~\ref{sec:prf_p3_cs}.

\subsection{Proving P4}

\begin{figure}[tbp]
    \centering
    \fbox{
        \begin{minipage}{0.95\linewidth}
            \vspace{0.5em}
            \centering
            \textbf{\underline{Protocol $\Pi_\text{P4}$}}
            \vspace{0.5em}

            \raggedright
                \textbf{Parameters:} A finite field $\mathbb{F}$. A clause encoding scheme $\text{Poly}_\phi$ and a literal encoding scheme $\phi$.
                
                \textbf{Input:} Both parties hold a commitment $[\Phi_\text{sec}]$ of all secret clauses to be checked. The number of clauses in $[\Phi_\text{sec}]$ is $n$. The two parties know the public literal set $\mathcal{L}_\text{pub}$ and the interface literal set $\mathcal{L}_\text{IO}$. Let $m$ denote the number of literals in $\mathcal{L}_\text{pub}\setminus\mathcal{L}_\text{IO}$.
                
                \textbf{Protocol:}
                \begin{enumerate}[leftmargin=*]
                    \item The prover sends $(\text{Commit},1)$ to $\mathcal{F}_\text{ZK}$. The two parties receive $[1]$ and send $(\text{Open},[1], 1)$ to $\mathcal{F}_\text{ZK}$. If the functionality aborts, the protocol aborts.
                    \item For every $i \in \{0,1,\cdots,n-1\}$ and every $j \in \{0,1,\cdots,m-1\}$, the verifier checks $c_i\in \Phi_\text{sec}$ with all disjoint public variables:
                    \begin{enumerate}
                        
                        \item The prover evaluates $\text{Poly}[c_i](l_j)$ and computes the multiplicative inverse $inv = \text{Poly}[c_i](l_j)^{-1}$ locally.
                        \item The prover sends $(\text{Commit}, inv)$ to $\mathcal{F}_\text{ZK}$. The two parties receive $[inv]$.
                        \item Both parties send $(\text{Eval},[c_i],l_j)$ to $\mathcal{F}_\text{ZK-Poly}$ and receive $[\text{Poly}[c_i](l_j)]$. 
                        \item Both parties send $(\text{Relation}, y=x_0\cdot x_1, [\text{Poly}[c_i](l_j)], [inv])$ to $\mathcal{F}_\text{ZK}$, and receive $[\text{Poly}[c_i](l_j)\cdot inv]$.
                        \item Both parties send $(\text{Equal},[\text{Poly}[c_i](l_j)\cdot inv],[1])$ to $\mathcal{F}_\text{ZK}$. If the functionality aborts, the protocol aborts.
                    \end{enumerate}
                    
                \end{enumerate}
                
        \end{minipage}
    }
    \caption{Protocol for proving P4.}
    \label{fig:prot_p4}
\end{figure}

\label{sec:non_inter}
To verify P4, the prover shows the verifier that the set of literals in $\Phi_{\text{sec}}$ is disjoint from the literals in $\Phi_{\text{pub}}$, except for the interface literals $\mathcal{L}_{\text{IO}}$. Based on the membership test defined in Eq.~\ref{eq:member}, a literal $l$ is not in clause $c$ if and only if $\text{Poly}[c](\phi(l)) \neq 0$. Hence the protocol verifies that for every literal $l \in \mathcal{L}_{\text{pub}} \setminus \mathcal{L}_{\text{IO}}$ and every clause $c \in \Phi_{\text{sec}}$, the polynomial evaluation $\text{Poly}[c](\phi(l))$ yields a non-zero value. We give the detailed protocol to prove P4 in Fig.\ref{fig:prot_p4}.

To prove that $\text{Poly}[c](\phi(l)) \neq 0$ without compromising zero-knowledge, we use a standard non-zero proof technique. The evaluation is performed through the \textbf{Eval} in $\mathcal{F}_\text{ZK-Poly}$, and the result is a commitment. Rather than opening the commitment, the prover locally calculates $inv = (\text{Poly}[c](\phi(l)))^{-1}$ and provides a commitment $[inv]$. The protocol then verifies that $[ \text{Poly}[c](\phi(l)) ] \cdot [inv] = [1]$. The existence of the multiplicative inverse in $\mathbb{F}$ guarantees that the evaluation is non-zero, while preventing any leakage of $\Phi_\text{sec}$.

\noindent
\textbf{Completeness.} 
When the two parties are honest, the completeness of the protocol $\Pi_\text{P4}$ is guaranteed by the completeness of the membership test in Eq.~\ref{eq:member} and the ZK backend.

\noindent
\textbf{Soundness.}
The soundness of the protocol $\Pi_\text{P4}$ is defined as: if $\exists l\in\mathcal{L}_\text{pub} \setminus \mathcal{L}_\text{IO}, l\in \Phi_\text{sec}$, the verifier will accept $\Pi_\text{P4}$ with negligible probability $\text{negl}(\lambda)$.
 Consider a malicious prover $\mathcal{P}^*$. The protocol assumes that both parties know $\mathcal{L}_\text{pub} \setminus \mathcal{L}_\text{IO}$. If there exists a literal $l \in \mathcal{L}_{\text{pub}} \setminus \mathcal{L}_{\text{IO}}$ such that $l \in \Phi_{\text{sec}}$, the only way for $\mathcal{P}^*$ to convince the verifier is to cheat in $\mathcal{F}_{\text{ZK-Poly}}$ and $\mathcal{F}_{\text{ZK}}$. However, the soundness of the ZK backend ensures that the verifier will accept the protocol with negligible probability $\text{negl}(\lambda)$.



\subsection{Putting Everything Together}
\label{sec:m_protocol}

We propose our main protocol $\Pi_\text{ZK-CEC}$ for zero-knowledge combinational equivalence checking in Fig.~\ref{fig:prot_zkcec} by composing $\Pi_\text{P1}$, $\Pi_\text{P2}$, $\Pi_\text{P3}$, and $\Pi_\text{P4}$ together. The soundness of the complete protocol relies on the sequential enforcement of the blueprint properties defined in Sec.~\ref{sec:blutprint}. While $\Pi_\text{P2}$ conditionally proves the unsatisfiability of the miter, it relies on $\Pi_\text{P3}$ and $\Pi_\text{P4}$ to guarantee that the underlying secret formula $\Phi_\text{sec}$ is satisfiable and isolated from the public property. This composition proves that the prover committed design $[\Phi_\text{impl}]$ is equivalent to the public specification $\Phi_\text{spec}$.

\begin{figure}[tbp]
    \centering
    \fbox{
        \begin{minipage}{0.95\linewidth}
            \vspace{0.5em}
            \centering
            \textbf{\underline{Protocol $\Pi_\text{ZK-CEC}$}}
            \vspace{0.5em}

            \raggedright
                \textbf{Parameters:} A finite field $\mathbb{F}$. A clause encoding scheme $\text{Poly}_\phi[\cdot]$ and a literal encoding scheme $\phi(\cdot)$.
                
                \textbf{Input:} The prover has a secret circuit $\Phi_\text{sec}=C_\text{impl}$. Both parties know a public specification $C_\text{spec}$.
                
                \textbf{Protocol:}
                \begin{enumerate}[leftmargin=*]
                    \item The prover builds the miter circuit, and transform $\Phi_\text{sec}$ from $C_\text{impl}$ and $\Phi_\text{pub}$ from $C_\text{spec}$, $C_\text{IO}$, and $l_\text{out}$.
                    \item The prover generates a secret key $k$ for the hash function $H$ in $\phi(\cdot)$.
                    
                    \item The prover encodes $\Phi_\text{sec}$ with $\text{Poly}_\phi[\cdot]$, and sends $(\text{Commit},\text{Poly}_\phi[\Phi_\text{sec}])$ to $\mathcal{F}_\text{ZK-Poly}$. The two parties receives $[\Phi_\text{sec}]$.
                    
                    \item The prover encodes $\Phi_\text{pub}$ with $\text{Poly}_\phi[\cdot]$, and sends $(\text{Commit},\text{Poly}_\phi[\Phi_\text{pub}])$ to $\mathcal{F}_\text{ZK-Poly}$. The two parties receives $[\Phi_\text{pub}]$.
                    \item The two parties run $\Pi_\text{P1}$ on input of $[\Phi_\text{pub}]$ and $C_\text{spec}$, and obtain $\mathcal{L}_\text{pub}$.
                    \item The prover locally solves $\Phi_\text{miter} = \Phi_\text{sec}\land \Phi_\text{pub}$ and obtains a refutation proof. The prover also knows the length of the refutation proof $R$ and the maximum width of all clauses $w$.
                    \item The two parties run $\Pi_\text{P2}$ on input of $[\Phi_\text{miter}]$, $R$, and $w$.
                    \item The prover locally solves $\Phi_\text{sec}$ and obtains an assignment $\omega$ that satisfies $\Phi_\text{sec}$.
                    \item The two parties run $\Pi_\text{P3}$ on input of $[\Phi_\text{sec}]$ and $\omega$.
                    \item The two parties run $\Pi_\text{P4}$ on input of $[\Phi_\text{sec}]$ and $\mathcal{L}_\text{pub}$.
                    
                \end{enumerate}
                
        \end{minipage}
    }
    \caption{Protocol for zero-knowledge combinational equivalence checking.}
    \label{fig:prot_zkcec}
\end{figure}

\begin{theorem}
\label{the:main}
$\Pi_\text{ZK-CEC}$ is a zero-knowledge proof of combinational equivalence.
\end{theorem}
We provide a proof sketch for Theorem~\ref{the:main} in Appendix~\ref{sec:prf_main}, which argues that the protocol satisfies completeness, soundness, and zero-knowledge.

\subsection{Practical Optimization with Extra Leakage}
\label{sec:opt}

To mitigate the computational overhead associated with large refutation proofs, we introduce a resolvent compression optimization designed to improve the performance of $\Pi_{P2}$. Notably, this optimization is also applicable to the original ZKUNSAT framework.

The insight is that only part of the resolvents would be used as antecedents multiple times. Those resolvents are \textit{learned clauses} in conflict-driven clause learning (CDCL) algorithm~\cite{marques2002grasp} that is widely used in modern SAT solvers. Therefore, we merge multiple intermediate resolution steps into a single derivation and store only those \textit{learned clauses} in the ZK-ROM, which reduces the overhead of permutation checks. We provide a simplified refutation proof structure example in Appendix~\ref{sec:prf_example} to give an intuitive understanding.

While the verifier remains unknown about the antecedents and resolvents used in a resolution step, this optimization introduces additional leakage of proof structure, where the verifier can determine the length of each sub-tree. To reconstruct the original formula, the verifier is required to first correctly guess the proof tree structure and then use it to recover the formula. 

We demonstrate that formula reconstruction remains sufficiently difficult under the optimization by giving a lower bound on the number of possible proof tree structures in the first step. For the sake of analysis, we assume each sub-tree is used only once and composed in the sequence they are checked, which only loosens the lower bound.
For a compressed resolution proof with $R'$ steps, let $n_j$ denote the number of possible antecedent positions at the $j$-th sub-tree. The number of potential structures for constructing the proof tree from the sub-trees is given by:
\begin{equation}
\label{eq:tree}
    N > \prod_{i=1}^{R'-1} \left( \sum_{j=i+1}^{R'} n_j \right) \geq (R'-1)!
\end{equation}
In the case of a sub-tree with a single resolution, the placement of a subsequent sub-tree occupies only one of the two antecedent positions. For sub-trees involving multiple resolutions, we assume that the subsequent sub-tree always occupies a possible position. Consequently, $n_j$ is defined as:
\begin{equation}
    n_j = 
        \begin{cases} 
        1 & \text{if } n_{\text{res}_j} = 1\\
        n_{\text{res}_j} - 1 & \text{if } n_{\text{res}_j} > 1, j \neq R'\\
        n_{\text{res}_j} & \text{if } n_{\text{res}_j} > 1, j = R';
        \end{cases}
\end{equation}
where $n_{\text{res}_j}$ denotes the number of resolvents within the $j$-th sub-tree. The term $(n_{\text{res}_j}-1)$ is because each preceding sub-tree ($j \neq R'$) occupies one possible position. We provide the step-by-step derivation of $N$ for the example in Appendix~\ref{sec:prf_example}.
The loose lower bound $N$ exceeds the factorial of $(R'-1)$, and $R'$ is large enough in practical design. For example, a 4-bit adder typically yields $R'=90$. Therefore, the factorial complexity renders the enumeration of potential proof structures sufficiently complex.

\subsection{Discussion on Fair Exchange}
In the context of the trustworthy 3PIP supply chain, ensuring the soundness of the verification process alone is insufficient; guaranteeing fair exchange between the vendor and the system integrator is equally important. While existing works allow vendors to present valid proofs~\cite{mouris2020pythia,mouris2022zk}, they lack a binding between the IP being transacted and the proof.

Our protocol addresses this issue through the commitment sche\-me in VOLE-based ZK, where the IP and the input commitments of the proof are bound. During the exchange, these commitments are opened and verified. By shifting the verification focus from design properties to key validity, our framework decouples the checking process from the design while reducing complexity. Furthermore, applying individual commitments to each clause within the formula enables gradual release during the transaction. By facilitating fair transactions through optimistic fair exchange~\cite{asokan1998optimistic,eckey2020optiswap} or smart contracts~\cite{zheng2020overview}, the two parties can achieve a fair trade.
\section{Implementation and Results}
\label{sec:impl}

\begin{table*}
\caption{Evaluation results.}
\label{tab:result}
\begin{tabular}{l|ccccSSSSS|cSS}
\toprule
\textbf{Circuit} & \textbf{Lits.} & \textbf{Cls.} & \textbf{$R$} & \textbf{$W$} & \textbf{$t_\text{P1}$ (s)} & \textbf{$t_\text{P2}$ (s)} & \textbf{$t_\text{P3}$ (ms)} & \textbf{$t_\text{P4}$ (ms)} & \textbf{$t_\text{Total}$ (s)} & \textbf{$R'$} & \textbf{$t'_\text{Total}$ (s)} & \textbf{Speedup($\times$)} \\
\midrule
Adder\_4x4 & 25 & 58 & 1225 & 22 & 0.010000 & 3.300000 & 0.410000 & 1.290000 & 3.310000 & 90 & 1.760000 & 1.880682 \\
Adder\_5x5 & 32 & 75 & 1980 & 22 & 0.020000 & 4.850000 & 0.650000 & 2.060000 & 4.880000 & 103 & 2.380000 & 2.050420 \\
Adder\_6x6 & 39 & 92 & 1938 & 18 & 0.020000 & 4.330000 & 0.820000 & 2.850000 & 4.350000 & 138 & 2.160000 & 2.013889 \\
Adder\_7x7 & 47 & 111 & 1500 & 19 & 0.010000 & 3.780000 & 1.240000 & 4.140000 & 3.800000 & 136 & 1.970000 & 1.928934 \\
Adder\_8x8 & 54 & 128 & 1671 & 17 & 0.020000 & 4.010000 & 1.490000 & 5.100000 & 4.040000 & 145 & 2.050000 & 1.970732 \\
Adder\_9x9 & 61 & 145 & 2272 & 17 & 0.020000 & 5.220000 & 2.110000 & 31.710000 & 5.270000 & 183 & 2.540000 & 2.074803 \\
Adder\_10x10 & 68 & 162 & 2565 & 19 & 0.020000 & 5.760000 & 2.720000 & 67.240000 & 5.850000 & 207 & 2.940000 & 1.989796 \\
Comp\_6x6 & 53 & 121 & 2803 & 23 & 0.050000 & 6.610000 & 1.740000 & 3.870000 & 6.660000 & 163 & 2.940000 & 2.265306 \\
Comp\_7x7 & 59 & 135 & 3850 & 29 & 0.080000 & 9.420000 & 2.610000 & 19.350000 & 9.520000 & 212 & 4.220000 & 2.255924 \\
Comp\_8x8 & 72 & 168 & 4770 & 31 & 0.140000 & 11.870000 & 4.350000 & 8.550000 & 12.020000 & 220 & 5.210000 & 2.307102 \\
Comp\_9x9 & 70 & 156 & 3777 & 26 & 0.080000 & 8.750000 & 3.430000 & 27.440000 & 8.870000 & 226 & 3.870000 & 2.291990 \\
Comp\_10x10 & 82 & 86 & 5570 & 33 & 0.210000 & 14.200000 & 5.760000 & 12.390000 & 14.430000 & 298 & 6.030000 & 2.393035 \\
Mul\_fp8 & 130 & 354 & 10828 & 33 & 0.320000 & 26.660000 & 16.720000 & 281.250000 & 27.270000 & 373 & 11.440000 & 2.383741 \\
Mul\_2x2 & 12 & 24 & 199 & 14 & 0.000000 & 1.220000 & 0.070000 & 0.250000 & 1.220000 & 34 & 1.160000 & 1.051724 \\
Mul\_3x3 & 37 & 101 & 1970 & 25 & 0.020000 & 4.980000 & 1.090000 & 2.880000 & 5.010000 & 97 & 2.420000 & 2.070248 \\
Mul\_4x4 & 84 & 246 & 16250 & 50 & 0.830000 & 52.000000 & 10.900000 & 136.990000 & 52.970000 & 412 & 20.130000 & 2.631396 \\
Mul\_5x5 & 144 & 436 & 100619 & 72 & 9.060000 & 414.480000 & 46.440000 & 506.480000 & 424.090000 & 1640 & 159.850000 & 2.653050 \\
Mul\_6x6 & 222 & 683 & 667234 & 106 & 112.870000 & 4010.250000 & 282.750000 & 1376.320000 & 4124.780000 & 7253 & 1454.740000 & 2.835407 \\
\midrule
Arbiter\_4 & 17 & 21 & 97 & 12 & 0.000000 & 1.160000 & 0.080000 & 0.190000 & 1.160000 & 27 & 1.000000 & 1.160000 \\
Barrel\_shifter\_4 & 26 & 58 & 879 & 19 & 0.010000 & 2.570000 & 0.380000 & 1.030000 & 2.580000 & 72 & 1.550000 & 1.664516 \\
Barrel\_shifter\_8 & 72 & 180 & 1891 & 39 & 0.100000 & 6.250000 & 5.600000 & 56.410000 & 6.410000 & 94 & 2.800000 & 2.289286 \\
CRC\_8 & 37 & 84 & 3304 & 24 & 0.050000 & 7.620000 & 0.890000 & 2.380000 & 7.670000 & 231 & 3.420000 & 2.242690 \\
CRC\_16 & 87 & 206 & 16742 & 36 & 0.540000 & 42.230000 & 7.190000 & 71.580000 & 42.850000 & 938 & 17.970000 & 2.384530 \\
Decoder\_3to8 & 19 & 47 & 443 & 22 & 0.010000 & 1.700000 & 0.270000 & 1.040000 & 1.700000 & 37 & 1.300000 & 1.307692 \\
Decoder\_4to16 & 31 & 78 & 1615 & 38 & 0.040000 & 5.200000 & 1.040000 & 4.370000 & 5.240000 & 71 & 2.490000 & 2.104418 \\
Decoder\_5to32 & 58 & 157 & 5842 & 72 & 0.620000 & 26.150000 & 6.870000 & 134.100000 & 26.910000 & 139 & 10.340000 & 2.602515 \\
FSM & 20 & 48 & 229 & 12 & 0.000000 & 1.290000 & 0.170000 & 0.470000 & 1.290000 & 31 & 1.130000 & 1.141593 \\
Parity\_8 & 15 & 28 & 330 & 11 & 0.000000 & 1.430000 & 0.090000 & 0.140000 & 1.430000 & 53 & 1.190000 & 1.201681 \\
Encoder\_8 & 26 & 51 & 326 & 16 & 0.000000 & 1.430000 & 0.290000 & 0.900000 & 1.440000 & 58 & 1.200000 & 1.200000 \\
Encoder\_16 & 67 & 147 & 1145 & 24 & 0.010000 & 3.340000 & 2.940000 & 21.800000 & 3.380000 & 144 & 1.850000 & 1.827027 \\
Voting\_9 & 52 & 140 & 3332 & 24 & 0.060000 & 7.510000 & 2.090000 & 3.460000 & 7.570000 & 186 & 3.330000 & 2.273273 \\
Voting\_17 & 103 & 282 & 60565 & 40 & 2.160000 & 159.630000 & 12.690000 & 101.170000 & 161.910000 & 2063 & 63.510000 & 2.549362 \\
\midrule
GFMul\_4x4 & 42 & 120 & 10441 & 30 & 0.240000 & 23.900000 & 1.710000 & 3.630000 & 24.140000 & 392 & 9.720000 & 2.483539 \\
GFMul\_8x8 & 169 & 538 & 1582909 & 109 & 284.760000 & 9691.580000 & 98.720000 & 896.820000 & 9977.330000 & 30303 & 3646.550000 & 2.736101 \\
S-box\_AES & 211 & 700 & 236862 & 298 & 275.600000 & 4660.740000 & 720.310000 & 14298.100000 & 4951.360000 & 1754 & 1763.090000 & 2.808342 \\
S-box\_ASCON & 24 & 68 & 3159 & 35 & 0.090000 & 9.210000 & 0.650000 & 43.440000 & 9.340000 & 120 & 3.770000 & 2.477454 \\
S-box\_PRESENT & 57 & 155 & 1476 & 29 & 0.020000 & 4.600000 & 2.940000 & 29.130000 & 4.660000 & 98 & 2.240000 & 2.080357 \\
S-box\_SM4 & 124 & 403 & 188098 & 300 & 228.910000 & 3770.040000 & 144.120000 & 7727.020000 & 4006.820000 & 1420 & 1450.960000 & 2.761496 \\
\bottomrule
\end{tabular}
\end{table*}

\subsection{Implementation}

We implement our protocol $\Pi_\text{ZK-CEC}$ based on the open-source code of ZKUNSAT~\cite{luo2022proving}. We use the EMP-ZK library of the EMP-toolkit~\cite{emp_toolkit} as our VOLE-based ZK backend. We use the NTL library~\cite{shoup2001ntl} for polynomial arithmetic over finite fields. The finite field instantiated in our implementation is $\mathbb{F}_{128}$. We use libsodium~\cite{Denis_libsodium} for generating the key $k$ and implementing BLAKE2b~\cite{rfc7693} as the keyed hash function in our protocol, where BLAKE2b can serve directly as a PRF. The largest bitwidth of literals $w_\text{lit}$ is set to be 64. We represent the clause indices as 24-bit integers, which supports refutation proof with at most $2^{23}$ resolutions. Additionally, we make some optimizations to the original ZKUNSAT code. 

\subsection{Evaluation Setup}

\begin{figure*}[tbp]
\centerline{\includegraphics[width=\textwidth]{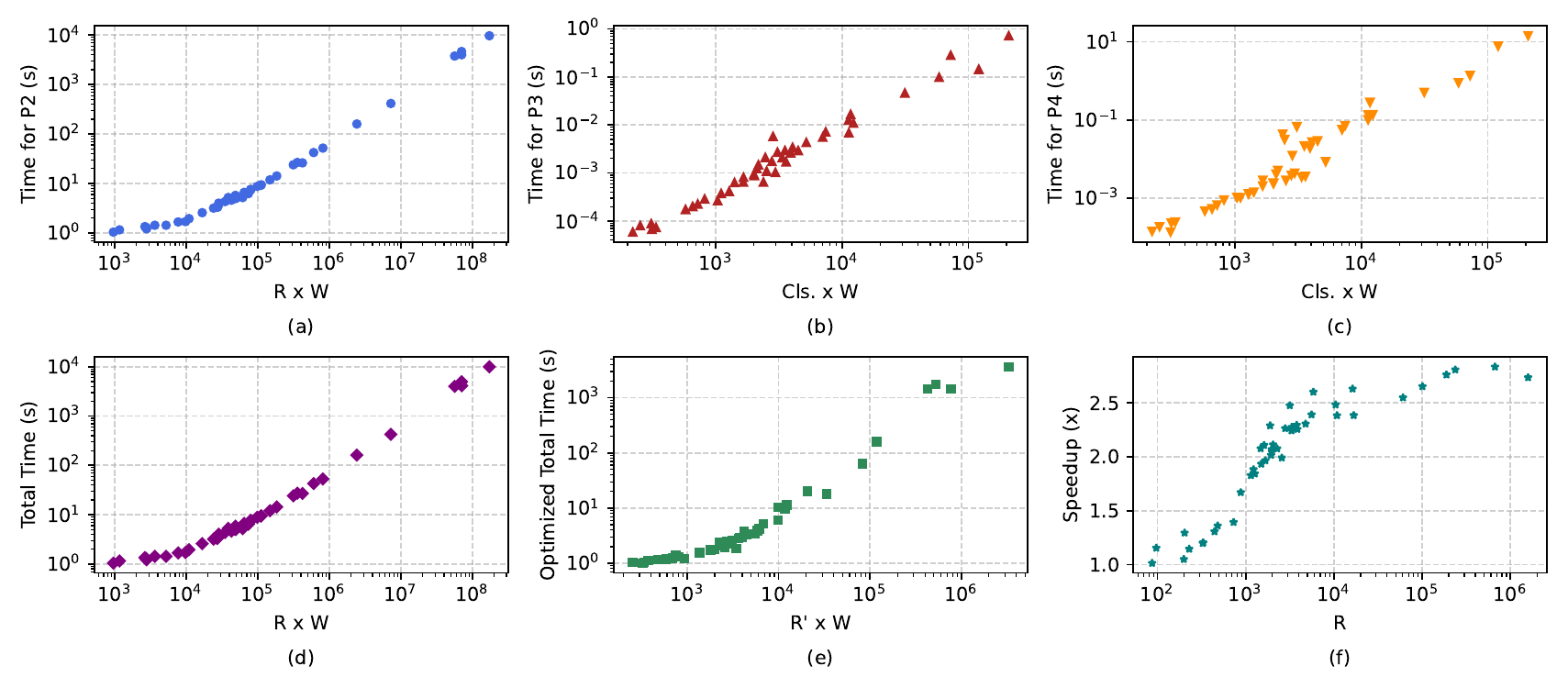}}
\caption{Proving time vs. input size: (a) Proving time or P2 (b) Proving time for P3 (c) Proving time for P4 (d) Total proving time (e) Total proving time after optimization (f) Speedup in the total proving time after optimization.}
\label{fig:perf}
\end{figure*}

We evaluate ZK-CEC on a cloud computing node that equips a Lenovo ThinkSystem SR645 server with an AMD Rome 7H12 processor. In our experimental configuration, we run the prover and the verifier on a single node and allocate 80GB RAM for each party. The prover and verifier communicate via the local network. As noted in~\cite{luo2022proving,luick2024zksmt}, the bandwidth requirement between the two parties is approximately 100 Mbps; therefore, network communication does not constitute a performance bottleneck in our evaluation.

A set of 37 combinational circuits is used as the benchmark of our evaluation. These include arithmetic operators (adders, multipliers, comparators, and floating-point multipliers) and control logic such as barrel shifters, decoders, priority encoders, FSM transition logic, and arbiters. We also include error-detection modules (parity and cyclic redundancy checks) and cryptographic components, which involve finite field multipliers and S-Boxes for AES, SM4, Ascon, and Present. Finally, voting circuits are included as a common logic in multi-party computation.

We use Synopsys Design Compiler to synthesize the benchmark into gate-level netlists. For each circuit, we manually constructed a miter circuit by combining the netlist with its specification, which is then transformed into a CNF formula. We use PicoSAT~\cite{biere2008picosat} to solve the formula and generate the refutation proof.

\subsection{Evaluation Results}

Table~\ref{tab:result} presents the evaluation results for the benchmark circuits. In the table, Lits. and Cls. denote the number of literals and clauses, respectively, in the CNF formula derived from the implementation circuit. $R$ and $W$ represent the total number of stored resolvents (i.e., the proof length) and the padded width of the encoded polynomials. The execution time for verifying each property is denoted by $t_\text{Pi}$. In our implementation, the prover commits the resolvent clauses during the execution of $\Pi_{P1}$; therefore, the running times for $\Pi_\text{P1}$ include the time of commitment. We also implement and evaluate the optimization in Sec.~\ref{sec:opt}. $R'$ and $t'_{Total}$ denote the compressed proof length and the total execution time under the optimization, and the last column provides the speedup over the baseline after optimization.

As shown in Table~\ref{tab:result}, the verification of the UNSAT ($\Pi_\text{P2}$) accounts for the majority of the execution time. This observation is consistent with the fact that UNSAT is a co-NP-complete problem; the length of the resolution refutation $R$ can grow exponentially relative to the size of the initial formula $\Phi$. In contrast, the verification costs for $\Pi_\text{P3}$ (i.e. proving SAT) and $\Pi_\text{P4}$ scale linearly with the size of $\Phi_\text{impl}$. As the scale of the design increases, the refutation becomes the dominant performance bottleneck.

Fig.~\ref{fig:perf}~(a) illustrates the correlation between $t_\text{P2}$ and the refutation proof size. Fig.~\ref{fig:perf}~(b) and \ref{fig:perf}~(c) depict the proving times for $\Pi_\text{P3}$ and $\Pi_\text{P4}$ as a function of the size of $\Phi_\text{impl}$, respectively. Given that proving UNSAT is dominant, Fig.~\ref{fig:perf}~(d) compares the total execution time against the refutation proof size. 
These four time metrics exhibit a near-linear relationship on a log-log scale. Notably, as we scaled our evaluation to more complex combinational logic designs, we inevitably encountered resource constraints. The explosive growth of the refutation proofs for larger circuits led to out-of-memory and time-outs, which are the current scalability limits of our framework. We further evaluate our protocol with optimization. Fig.~\ref{fig:perf}~(e) illustrates the correlation between total execution time and the compressed proof size, while Fig.~\ref{fig:perf}~(f) depicts the speedup compared to the baseline. Notably, the speedup grows with the scale of design.

\section{Limitations and Discussion}

While ZK-CEC represents a foundational step toward zero-knowledge formal hardware verification, its practical deployment is currently constrained by the inherent scalability limits of both formal equivalence checking and the underlying zero-knowledge backend. Although the proposed framework imposes no theoretical bounds on circuit size, the required resolution refutation proofs can grow exponentially as the scale and complexity of the implementation circuit increase. This exponential growth translates directly into significant memory consumption and extended proving times. Additionally, the cryptographic operations required by the zero-knowledge framework introduce substantial computational overhead. As demonstrated in Table~\ref{tab:Runtime}, ZK-CEC incurs a roughly $2500\times$ overhead when verifying 4-bit and 5-bit multipliers compared to plaintext SAT solving. Together, these factors dictate the scalability limits of our current implementation.

To address these limitations and enable the verification of larger hardware IPs, future deployments of ZK-CEC can adopt a modular verification strategy. Following standard practices in electronic design automation (EDA), complex designs can be partitioned into smaller, more manageable modules for independent verification, thereby drastically reducing overall proof complexity~\cite{grumberg1994model,mukherjee2015equivalence}. Furthermore, this modular approach could naturally extend the capabilities of the ZK-CEC framework to verify sequential circuits by checking the combinational logic between pipeline registers.

\begin{table}[tbp]
  \centering
  \caption{Runtime Comparison between ZKCEC and SAT Solving}
  \label{tab:Runtime}
  \begin{tabular}{l|lll}
    \toprule
    \textbf{Design} & \textbf{SAT Solver} & \textbf{ZKCEC} & \textbf{Overhead} \\
    \midrule
    Mult\_4x4 & 0.0082 s & 20.13 s & $2454.88\times$\\
    Mult\_5x5 & 0.0636 s & 159.55 s & $2508.65\times$ \\
    \bottomrule
  \end{tabular}
\end{table}
\section{Related Work}
\label{sec:related}

\noindent
\textbf{Privacy Preserving Hardware Verification.}
The line of research most closely aligned with our work is privacy-preserving hardware verification. Notably, Pythia~\cite{mouris2020pythia} encodes the vendor's circuit and puts it into a zkVM as witness input. This framework allows the verifier to perform evaluations using private stimuli and prove output correctness, thereby enabling simulation-based verification between IP vendors and buyers in zero-knowledge. Building on simulation, zk-Sherlock~\cite{mouris2022zk} calculates the toggle rates of signals in the netlist within a zkVM, and the verifier can analyze the switching activity to ensure that the design is free of hardware Trojan triggers. Leveraging multi-party computation and logic locking,~\cite{mouris2023text} realizes privacy-preserving IP evaluation with high scalability and performance. Romeo~\cite{gouert2020romeo} and~\cite{konstantinou2015privacy} rely on HE for privacy-preserving IP verification. 
Simulation-based verification often struggles to catch corner cases. Our work is the first zero-knowledge formal hardware verification framework, providing a proof of functional equivalence between the IP and the specification.

\noindent
\textbf{Formal Methods in Zero-Knowledge.}
Recent advancements have introduced specialized arithmetization techniques to prove logic properties within zero-knowledge frameworks. 
ZK-SMT~\cite{luick2024zksmt} enables the validation of SMT formulas in zero-knowledge based on a set of inference rules, while zkPi~\cite{laufer2024zkpi} provides a zkSNARK for proofs in Lean, capable of expressing virtually all systems of formal reasoning. Other works focus on specific applications: Cheesecloth~\cite{cuellar2023cheesecloth} is a statement compiler that proves program vulnerabilities, and Crepe~\cite{kolesar2025coinductive} proves regular expression equivalence (a PSPACE-complete problem). However, these existing approaches to formal reasoning operate exclusively on public inputs, hiding only the proof witness. Although the authors of~\cite{luo2022proving} state secret program analysis as a potential direction, they do not implement it due to unresolved consistency challenges. Our work presents the first framework for the formal verification of private inputs, leveraging the underlying structure of the formula.


\noindent
\textbf{VOLE-based Zero-Knowledge Proof.}
VOLE-based zero-knowledge proof protocols have been a line of research in recent years~\cite{weng2021wolverine,yang2021quicksilver,baum2021mac,weng2021mystique,dittmer2020line}. Wolverine~\cite{weng2021wolverine} first proposes a ZKP system that enables an efficient prover in both running time and memory usage. LPZK~\cite{dittmer2020line} focuses on utilizing the linear relationships of VOLE-based commitments. QuickSilver~\cite{yang2021quicksilver} is a hybrid approach of Wolverine and LPZK, which supports both large and small fields while avoiding cut-and-choose. Mystique~\cite{weng2021mystique} allows the conversions between arithmetic and Boolean commitments and improves the matrix multiplication operation in the protocol, which enables a zero-knowledge proof of high-precision deep neural network inference. 

\section{Conclusion}
\label{sec:conclusion}

This paper proposes the first privacy-preserving hardware formal verification protocol, ZK-CEC, for proving circuit functional equivalence without revealing the structure of the secret design, which aims to build trust between IP vendors and buyers. The proposed main protocol $\Pi_\text{ZK-CEC}$ achieves completeness, soundness, and zero-knowledge. In the paper, we first analyze the soundness issue of the existing ZK framework on the secret formula verifying problem. Next, we propose a general blueprint for privacy-preserving property checking. Based on our blueprint, we design four sub-protocols and compose them into our main protocol. Finally, we implement and evaluate our protocol upon a diverse set of common circuits, which demonstrates that our blueprint only brings little overhead to the existing ZK framework. For future work, the proposed blueprint can be applied to other verification scenarios, such as software and cyber-physical systems. Future research may also explore leveraging LLMs to translate text-based specs into formulas~\cite{cosler2023nl2spec} for formal verification through our proposed blueprint.
\section*{Ethical Considerations}
This research addresses the fundamental challenge of establishing trust and enhancing reliability within the global hardware supply chain. Our work aims to empower system integrators to verify the functional correctness of secret designs without compromising the vendor's intellectual property. We strictly adhere to established ethical research guidelines. Consequently, we believe the methodologies and protocols presented in this paper do not lead to potentially negative outcomes, as they are designed to enhance security and facilitate fair transactions in the hardware supply chain.

\begin{acks}
Zunchen Huang and Chenglu Jin are (partially) supported by project CiCS of the research programme Gravitation, which is (partly) financed by the Dutch Research Council (NWO) under the grant 024.006.037. We thank SURF for providing computational resources. We thank anonymous reviewers for their constructive feedback. 
We thank Marten van Dijk and Ruzica Piskac for their insightful discussions regarding zero-knowledge proofs for formal verification. This paper was edited for grammar using Gemini.
\end{acks}

\bibliographystyle{ACM-Reference-Format}
\bibliography{main}

@inproceedings{luo2022proving,
  title={Proving UNSAT in zero knowledge},
  author={Luo, Ning and Antonopoulos, Timos and Harris, William R and Piskac, Ruzica and Tromer, Eran and Wang, Xiao},
  booktitle={Proceedings of the 2022 ACM SIGSAC Conference on Computer and Communications Security},
  pages={2203--2217},
  year={2022}
}

@inproceedings{mouris2020pythia,
  title={Pythia: Intellectual property verification in zero-knowledge},
  author={Mouris, Dimitris and Tsoutsos, Nektarios Georgios},
  booktitle={2020 57th ACM/IEEE Design Automation Conference (DAC)},
  pages={1--6},
  year={2020},
  organization={IEEE}
}

@inproceedings{mouris2022zk,
  title={Zk-sherlock: Exposing hardware trojans in zero-knowledge},
  author={Mouris, Dimitris and Gouert, Charles and Tsoutsos, Nektarios Georgios},
  booktitle={2022 IEEE Computer Society Annual Symposium on VLSI (ISVLSI)},
  pages={170--175},
  year={2022},
  organization={IEEE}
}

@inproceedings{mouris2023text,
  title={MPoC: Privacy-Preserving IP Verification Using Logic Locking and Secure Multiparty Computation},
  author={Mouris, Dimitris and Gouert, Charles and Tsoutsos, Nektarios Georgios},
  booktitle={2023 IEEE 29th International Symposium on On-Line Testing and Robust System Design (IOLTS)},
  pages={1--8},
  year={2023},
  organization={IEEE}
}

@inproceedings{gouert2020romeo,
  title={Romeo: Conversion and evaluation of HDL designs in the encrypted domain},
  author={Gouert, Charles and Tsoutsos, Nektarios Georgios},
  booktitle={2020 57th ACM/IEEE Design Automation Conference (DAC)},
  pages={1--6},
  year={2020},
  organization={IEEE}
}

@inproceedings{konstantinou2015privacy,
  title={Privacy-preserving functional IP verification utilizing fully homomorphic encryption},
  author={Konstantinou, Charalambos and Keliris, Anastasis and Maniatakos, Michail},
  booktitle={2015 Design, Automation \& Test in Europe Conference \& Exhibition (DATE)},
  pages={333--338},
  year={2015},
  organization={IEEE}
}

@inproceedings{luick2024zksmt,
  title={$\{$ZKSMT$\}$: A $\{$VM$\}$ for Proving $\{$SMT$\}$ Theorems in Zero Knowledge},
  author={Luick, Daniel and Kolesar, John C and Antonopoulos, Timos and Harris, William R and Parker, James and Piskac, Ruzica and Tromer, Eran and Wang, Xiao and Luo, Ning},
  booktitle={33rd USENIX Security Symposium (USENIX Security 24)},
  pages={3837--3845},
  year={2024}
}

@inproceedings{laufer2024zkpi,
  title={zkpi: Proving lean theorems in zero-knowledge},
  author={Laufer, Evan and Ozdemir, Alex and Boneh, Dan},
  booktitle={Proceedings of the 2024 on ACM SIGSAC Conference on Computer and Communications Security},
  pages={4301--4315},
  year={2024}
}

@article{kolesar2025coinductive,
  title={Coinductive Proofs of Regular Expression Equivalence in Zero Knowledge},
  author={Kolesar, John C and Ali, Shan and Antonopoulos, Timos and Piskac, Ruzica},
  journal={Proceedings of the ACM on Programming Languages},
  volume={9},
  number={OOPSLA2},
  pages={357--385},
  year={2025},
  publisher={ACM New York, NY, USA}
}

@inproceedings{cuellar2023cheesecloth,
  title={Cheesecloth:$\{$Zero-Knowledge$\}$ proofs of real world vulnerabilities},
  author={Cu{\'e}llar, Santiago and Harris, Bill and Parker, James and Pernsteiner, Stuart and Tromer, Eran},
  booktitle={32nd USENIX Security Symposium (USENIX Security 23)},
  pages={6525--6540},
  year={2023}
}

@article{davis1960computing,
  title={A computing procedure for quantification theory},
  author={Davis, Martin and Putnam, Hilary},
  journal={Journal of the ACM (JACM)},
  volume={7},
  number={3},
  pages={201--215},
  year={1960},
  publisher={ACM New York, NY, USA}
}

@article{robinson1965machine,
  title={A machine-oriented logic based on the resolution principle},
  author={Robinson, John Alan},
  journal={Journal of the ACM (JACM)},
  volume={12},
  number={1},
  pages={23--41},
  year={1965},
  publisher={ACM New York, NY, USA}
}

@inproceedings{sun2025committed,
  title={Committed Vector Oblivious Linear Evaluation and Its Applications},
  author={Sun, Yunqing and Liu, Hanlin and Yang, Kang and Yu, Yu and Wang, Xiao and Weng, Chenkai},
  booktitle={Proceedings of the 2025 ACM SIGSAC Conference on Computer and Communications Security},
  pages={3635--3648},
  year={2025}
}

@inproceedings{weng2021wolverine,
  title={Wolverine: Fast, scalable, and communication-efficient zero-knowledge proofs for boolean and arithmetic circuits},
  author={Weng, Chenkai and Yang, Kang and Katz, Jonathan and Wang, Xiao},
  booktitle={2021 IEEE Symposium on Security and Privacy (SP)},
  pages={1074--1091},
  year={2021},
  organization={IEEE}
}

@inproceedings{yang2021quicksilver,
  title={Quicksilver: Efficient and affordable zero-knowledge proofs for circuits and polynomials over any field},
  author={Yang, Kang and Sarkar, Pratik and Weng, Chenkai and Wang, Xiao},
  booktitle={Proceedings of the 2021 ACM SIGSAC Conference on Computer and Communications Security},
  pages={2986--3001},
  year={2021}
}

@inproceedings{baum2021mac,
  title={Mac n Cheese: zero-knowledge proofs for boolean and arithmetic circuits with nested disjunctions},
  author={Baum, Carsten and Malozemoff, Alex J and Rosen, Marc B and Scholl, Peter},
  booktitle={Annual International Cryptology Conference},
  pages={92--122},
  year={2021},
  organization={Springer}
}

@inproceedings{weng2021mystique,
  title={Mystique: Efficient conversions for $\{$Zero-Knowledge$\}$ proofs with applications to machine learning},
  author={Weng, Chenkai and Yang, Kang and Xie, Xiang and Katz, Jonathan and Wang, Xiao},
  booktitle={30th USENIX Security Symposium (USENIX Security 21)},
  pages={501--518},
  year={2021}
}

@article{dittmer2020line,
  title={Line-point zero knowledge and its applications},
  author={Dittmer, Samuel and Ishai, Yuval and Ostrovsky, Rafail},
  journal={Cryptology ePrint Archive},
  year={2020}
}

@incollection{tseitin1983complexity,
  title={On the complexity of derivation in propositional calculus},
  author={Tseitin, Grigori S},
  booktitle={Automation of reasoning: 2: Classical papers on computational logic 1967--1970},
  pages={466--483},
  year={1983},
  publisher={Springer}
}

@article{perrucci2007real,
  title={Real roots of univariate polynomials and straight line programs},
  author={Perrucci, Daniel and Sabia, Juan},
  journal={Journal of Discrete Algorithms},
  volume={5},
  number={3},
  pages={471--478},
  year={2007},
  publisher={Elsevier}
}

@inproceedings{franzese2021constant,
  title={Constant-overhead zero-knowledge for RAM programs},
  author={Franzese, Nicholas and Katz, Jonathan and Lu, Steve and Ostrovsky, Rafail and Wang, Xiao and Weng, Chenkai},
  booktitle={Proceedings of the 2021 ACM SIGSAC Conference on Computer and Communications Security},
  pages={178--191},
  year={2021}
}

@misc{emp_toolkit,
      author = "Xiao Wang and Alex J. Malozemoff and Jonathan Katz",
      title = {{EMP-toolkit: Efficient MultiParty computation toolkit}},
      howpublished = {\url{https://github.com/emp-toolkit}},
      year={2016}
}

@misc{shoup2001ntl,
  title={NTL: A library for doing number theory},
  author={Shoup, Victor and others},
  year={2001}
}

@software{Denis_libsodium,
author = {Denis, Frank},
license = {ISC},
title = {{libsodium}},
url = {https://github.com/jedisct1/libsodium}
}

@misc{rfc7693,
    series =    {Request for Comments},
    number =    7693,
    howpublished =  {RFC 7693},
    publisher = {RFC Editor},
    doi =       {10.17487/RFC7693},
    url =       {https://www.rfc-editor.org/info/rfc7693},
    author =    {Markku-Juhani O. Saarinen and Jean-Philippe Aumasson},
    title =     {{The BLAKE2 Cryptographic Hash and Message Authentication Code (MAC)}},
    pagetotal = 30,
    year =      2015,
    month =     nov,
}

@article{biere2008picosat,
  title={PicoSAT essentials},
  author={Biere, Armin},
  journal={Journal on Satisfiability, Boolean Modelling and Computation},
  volume={4},
  number={2-4},
  pages={75--97},
  year={2008},
  publisher={SAGE Publications Sage UK: London, England}
}

@inproceedings{ray2018protecting,
  title={Protecting the supply chain for automotives and IoTs},
  author={Ray, Sandip and Chen, Wen and Cammarota, Rosario},
  booktitle={Proceedings of the 55th Annual Design Automation Conference},
  pages={1--4},
  year={2018}
}

@article{saleh2006system,
  title={System-on-chip: Reuse and integration},
  author={Saleh, Resve and Wilton, Steve and Mirabbasi, Shahriar and Hu, Alan and Greenstreet, Mark and Lemieux, Guy and Pande, Partha Pratim and Grecu, Cristian and Ivanov, Andre},
  journal={Proceedings of the IEEE},
  volume={94},
  number={6},
  pages={1050--1069},
  year={2006},
  publisher={IEEE}
}

@inproceedings{goldberg2001using,
  title={Using SAT for combinational equivalence checking},
  author={Goldberg, Evguenii I and Prasad, Mukul R and Brayton, Robert K},
  booktitle={Proceedings Design, Automation and Test in Europe. Conference and Exhibition 2001},
  pages={114--121},
  year={2001},
  organization={IEEE}
}

@article{baumgarten2011case,
  title={A case study in hardware Trojan design and implementation},
  author={Baumgarten, Alex and Steffen, Michael and Clausman, Matthew and Zambreno, Joseph},
  journal={International Journal of Information Security},
  volume={10},
  number={1},
  pages={1--14},
  year={2011},
  publisher={Springer}
}

@inproceedings{waksman2011silencing,
  title={Silencing hardware backdoors},
  author={Waksman, Adam and Sethumadhavan, Simha},
  booktitle={2011 IEEE Symposium on Security and Privacy},
  pages={49--63},
  year={2011},
  organization={IEEE}
}

@inproceedings{rad2008power,
  title={Power supply signal calibration techniques for improving detection resolution to hardware Trojans},
  author={Rad, Reza M and Wang, Xiaoxiao and Tehranipoor, Mohammad and Plusquellic, Jim},
  booktitle={2008 IEEE/ACM International Conference on Computer-Aided Design},
  pages={632--639},
  year={2008},
  organization={IEEE}
}

@inproceedings{jin2008hardware,
  title={Hardware Trojan detection using path delay fingerprint},
  author={Jin, Yier and Makris, Yiorgos},
  booktitle={2008 IEEE International workshop on hardware-oriented security and trust},
  pages={51--57},
  year={2008},
  organization={IEEE}
}

@article{rostami2014primer,
  title={A primer on hardware security: Models, methods, and metrics},
  author={Rostami, Masoud and Koushanfar, Farinaz and Karri, Ramesh},
  journal={Proceedings of the IEEE},
  volume={102},
  number={8},
  pages={1283--1295},
  year={2014},
  publisher={IEEE}
}

@article{love2011proof,
  title={Proof-carrying hardware intellectual property: A pathway to trusted module acquisition},
  author={Love, Eric and Jin, Yier and Makris, Yiorgos},
  journal={IEEE Transactions on Information Forensics and Security},
  volume={7},
  number={1},
  pages={25--40},
  year={2011},
  publisher={IEEE}
}

@inproceedings{jin2012proof,
  title={Proof carrying-based information flow tracking for data secrecy protection and hardware trust},
  author={Jin, Yier and Makris, Yiorgos},
  booktitle={2012 IEEE 30th VLSI Test Symposium (VTS)},
  pages={252--257},
  year={2012},
  organization={IEEE}
}

@book{kundu2018network,
  title={Network-on-chip: the next generation of system-on-chip integration},
  author={Kundu, Santanu and Chattopadhyay, Santanu},
  year={2018},
  publisher={Taylor \& Francis}
}

@article{haider2019advancing,
  title={Advancing the state-of-the-art in hardware trojans detection},
  author={Haider, Syed Kamran and Jin, Chenglu and Ahmad, Masab and Shila, Devu Manikantan and Khan, Omer and van Dijk, Marten},
  journal={IEEE Transactions on Dependable and Secure Computing},
  volume={16},
  number={1},
  pages={18--32},
  year={2019},
  publisher={IEEE}
}

@inproceedings{haider2017advancing,
  title={Advancing the state-of-the-art in hardware trojans design},
  author={Haider, Syed Kamran and Jin, Chenglu and van Dijk, Marten},
  booktitle={2017 IEEE 60th International Midwest Symposium on Circuits and Systems (MWSCAS)},
  pages={823--826},
  year={2017},
  organization={IEEE}
}

@inproceedings{asokan1998optimistic,
  title={Optimistic fair exchange of digital signatures},
  author={Asokan, Nadarajah and Shoup, Victor and Waidner, Michael},
  booktitle={International Conference on the Theory and Applications of Cryptographic Techniques},
  pages={591--606},
  year={1998},
  organization={Springer}
}

@inproceedings{eckey2020optiswap,
  title={Optiswap: Fast optimistic fair exchange},
  author={Eckey, Lisa and Faust, Sebastian and Schlosser, Benjamin},
  booktitle={Proceedings of the 15th ACM Asia Conference on Computer and Communications Security},
  pages={543--557},
  year={2020}
}

@article{zheng2020overview,
  title={An overview on smart contracts: Challenges, advances and platforms},
  author={Zheng, Zibin and Xie, Shaoan and Dai, Hong-Ning and Chen, Weili and Chen, Xiangping and Weng, Jian and Imran, Muhammad},
  journal={Future Generation Computer Systems},
  volume={105},
  pages={475--491},
  year={2020},
  publisher={Elsevier}
}

@inproceedings{nielsen2012new,
  title={A new approach to practical active-secure two-party computation},
  author={Nielsen, Jesper Buus and Nordholt, Peter Sebastian and Orlandi, Claudio and Burra, Sai Sheshank},
  booktitle={Annual Cryptology Conference},
  pages={681--700},
  year={2012},
  organization={Springer}
}

@article{marques2002grasp,
  title={GRASP: A search algorithm for propositional satisfiability},
  author={Marques-Silva, Joao P and Sakallah, Karem A},
  journal={IEEE Transactions on Computers},
  volume={48},
  number={5},
  pages={506--521},
  year={2002},
  publisher={IEEE}
}

@article{grumberg1994model,
  title={Model checking and modular verification},
  author={Grumberg, Orna and Long, David E},
  journal={ACM Transactions on Programming Languages and Systems (TOPLAS)},
  volume={16},
  number={3},
  pages={843--871},
  year={1994},
  publisher={ACM New York, NY, USA}
}

@inproceedings{mukherjee2015equivalence,
  title={Equivalence checking using trace partitioning},
  author={Mukherjee, Rajdeep and Kroening, Daniel and Melham, Tom and Srivas, Mandayam},
  booktitle={2015 IEEE Computer Society Annual Symposium on VLSI},
  pages={13--18},
  year={2015},
  organization={IEEE}
}

@inproceedings{hitchcock2015np,
  title={On the NP-completeness of the minimum circuit size problem},
  author={Hitchcock, John M and Pavan, Aduri},
  booktitle={35th IARCS Annual Conference on Foundations of Software Technology and Theoretical Computer Science (FSTTCS 2015)},
  pages={236--245},
  year={2015},
  organization={Schloss Dagstuhl--Leibniz-Zentrum f{\"u}r Informatik}
}

@article{reyhani2018smashing,
  title={Smashing the implementation records of AES S-box},
  author={Reyhani-Masoleh, Arash and Taha, Mostafa and Ashmawy, Doaa},
  journal={IACR transactions on cryptographic hardware and embedded systems},
  pages={298--336},
  year={2018}
}

@article{maximov2019new,
  title={New circuit minimization techniques for smaller and faster AES SBoxes},
  author={Maximov, Alexander and Ekdahl, Patrik},
  journal={IACR Transactions on Cryptographic Hardware and Embedded Systems},
  pages={91--125},
  year={2019}
}

@inproceedings{cosler2023nl2spec,
  title={nl2spec: Interactively translating unstructured natural language to temporal logics with large language models},
  author={Cosler, Matthias and Hahn, Christopher and Mendoza, Daniel and Schmitt, Frederik and Trippel, Caroline},
  booktitle={International Conference on Computer Aided Verification},
  pages={383--396},
  year={2023},
  organization={Springer}
}

@article{karthikeyan2025towards,
  title={Towards Practical Zero-Knowledge Proof for PSPACE},
  author={Karthikeyan, Ashwin and Liu, Hengyu and Meel, Kuldeep S and Luo, Ning},
  journal={arXiv preprint arXiv:2511.15071},
  year={2025}
}

\appendix

\section{Proofs}

\subsection{Proof of Lemma~\ref{lemma:sat}}
\label{sec:prf_lemma_sat}
We prove Lemma~\ref{lemma:sat} by contradiction. Given $\Phi_a \land \Phi_b$ is unsatisfiable, and two assignment sets that are not disjoint, then there must exist an assignment $\omega_s$ on $\mathcal{V}_s = \mathcal{V}_a \cap \mathcal{V}_b$, such that $\omega_s \in \{ \omega \mid \omega \models \exists \mathcal{V}'_a\Phi_a \} \cap \{ \omega \mid \omega \models \exists \mathcal{V}'_b\Phi_b \}$. By the definition of existential quantification, any model (i.e., assignment) of $\exists x\Phi$ can be lifted to a model of the original formula $\Phi$. Therefore, we lift $\omega_s$ to $\omega_a$ and  $\omega_b$ for $\Phi_a$ and $\Phi_b$, respectively. We define a new assignment $\omega_{union}$ over $\mathcal{V}_a\cup \mathcal{V}_b$ as:
\begin{equation*}
    \omega_{union}(x) = 
        \begin{cases} 
        \omega_s(x) & \text{if } x \in \mathcal{V}_a \cap \mathcal{V}_b\\
        \omega_a(x) & \text{if } x \in \mathcal{V}_a \setminus \mathcal{V}_s \\
        \omega_b(x) & \text{if } x \in \mathcal{V}_b \setminus \mathcal{V}_s
        \end{cases}
\end{equation*}
Since $\omega_{union}\models \Phi_a$ and $\omega_{union}\models \Phi_b$, we have $\omega_{union}\models \Phi_a\land\Phi_b$. This implies that $\Phi_a\land\Phi_b$ is satisfiable, which contradicts the initial assumption that $\Phi_a\land\Phi_b$ is unsatisfiable. This concludes the proof.

\subsection{Proof Sketch of $\Pi_\text{P3}$ Conditional Soundness}
\label{sec:prf_p3_cs}
Consider a malicious prover $\mathcal{P}^*$. If $\Phi_{\text{sec}}$ is unsatisfiable, there are two potential strategies for the prover to cheat in the protocol:
\begin{enumerate}[leftmargin=*]
\item Cheating within functionalities $\mathcal{F}_\text{ZK}$ and $\mathcal{F}_\text{ZK-Poly}$.
The prover may attempt to manipulate the protocol by modifying its own shares in the VOLE-based commitments, thereby binding the commitments to malicious values to force the verifier to accept. However, according to the proofs in~\cite{yang2021quicksilver}, as long as the verifier does not disclose its own commitment shares to $\mathcal{P}^*$, the probability that $\mathcal{F}_{ZK}$ and $\mathcal{F}_{ZK\text{-}poly}$ accept during the Equal and PoPIdt instructions, when they should abort, is $\text{negl}(\lambda)$.
\item Cheating on the assignment of $\Phi_{\text{sec}}$.
Assuming no cheating occurs within $\mathcal{F}_\text{ZK}$ and $\mathcal{F}_\text{ZK-Poly}$, the prover may attempt to provide a malicious assignment $\omega^*$ to convince the verifier. By the soundness of the resolution rule~\cite{robinson1965machine}, if a formula is satisfiable, any clause derived from it via resolution must also be satisfied by the same assignment. Since $\Phi_\text{sec}$ is unsatisfiable, there exists a resolution derivation that results in a contradiction, specifically two complementary literals $l$ and $\neg l$.
For $\omega^*$ to satisfy both, it must contain this pair of complementary literals.
If $\phi(l)$ and $\phi(\neg l)$ satisfy the encoding constraint in Eq.~\ref{eq:constraint}, the protocol will abort at Step (3) of $\Pi_\text{P3}$.
If $\phi(l)$ and $\phi(\neg l)$ do not satisfy this constraint, the inconsistency will be detected, and the preceding protocol $\Pi_{P2}$ will abort. Consequently, the probability of $\mathcal{P}^*$ successfully cheating in Case 2 is zero.
\end{enumerate}
Based on the analysis of the cases, when $\Phi_{\text{sec}}$ is unsatisfiable, a malicious prover has only a negligible probability of convincing the verifier that $\Phi_{\text{sec}}$ is satisfiable.

\subsection{Proof Sketch of Theorem~\ref{the:main}}
\label{sec:prf_main}
We now prove that the protocol $\Pi_\text{ZK-CEC}$ satisfies completeness, soundness, and zero-knowledge.

\noindent
\textbf{Completeness.} Completeness follows directly from the deterministic construction of the protocol.

\noindent
\textbf{Soundness.} First, the probability of $\mathcal{P}^*$ successfully cheating in the functionalities $\mathcal{F}_\text{ZK}$ and $\mathcal{F}_\text{ZK-Poly}$ is bounded by a negligible function $\text{negl}(\lambda)$. Therefore, the ZK backend ensures that any attempt to cheat in sub-protocols $\Pi_\text{P1}$ and $\Pi_\text{P4}$ will make the protocol abort with overwhelming probability. The soundness of the refutation in $\Pi_\text{P2}$ depends on the validity of the commitment $[\Phi_\text{sec}]$. According to Lemma~\ref{lemma:sat}, $\Pi_\text{P3}$ and $\Pi_\text{P4}$ guarantee that the statement $[\Phi_\text{sec}]$ in $\Pi_\text{P2}$ is valid and non-contradictory. This ensures that any derived contradiction originates from the interface $\mathcal{V}_\text{IO}$ rather than a forged internal conflict, thereby establishing the soundness of $\Pi_{P2}$. Regarding the conditional soundness in $\Pi_{P3}$, if the complementary literals do not satisfy the enforced constraints, the resolution process in $\Pi_{P2}$ will result in an abort. Since each component is either cryptographically secure or logically constrained by the sub-protocols, $\Pi_\text{ZK-CEC}$ satisfies the soundness property.

\noindent
\textbf{Zero-knowledge.} We construct a simulator $S$ that holds a circuit $C'_\text{impl}$, where $C'_\text{impl}\equiv C_\text{spec}$. Under our defined leakage model, $C'_\text{impl}$ is indistinguishable from the real implementation $C_\text{impl}$ with respect to refutation proof size and gate-count. Throughout the protocol execution, the verifier’s view is restricted to VOLE commitments and the abort signal. The VOLE-based ZK backend instantiated in $\Pi_\text{ZK-CEC}$ ensures that all commitments and polynomial evaluations achieve information-theoretic security~\cite{yang2021quicksilver}. Therefore, in the verifier's view, all commitments are indistinguishable from uniform random values. Consequently, the simulated interaction is indistinguishable from the real protocol execution, thereby satisfying the zero-knowledge property.

\section{A Simplified Example of the Structure of a Refutation Proof}
\label{sec:prf_example}

\begin{figure}[tbp]
\centerline{\includegraphics[width=6cm]{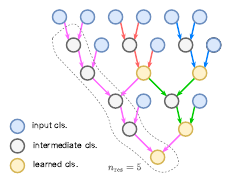}}
\caption{A refutation proof tree example.}
\label{fig:tree}
\end{figure}

Fig.~\ref{fig:tree} illustrates a simplified example of a refutation proof. The blue nodes represent the input clauses, while the yellow nodes denote the learned clauses; the remaining gray nodes represent intermediate clauses. Arrows of different colors correspond to different sub-tree blocks. Each sub-tree concludes by deriving a learned clause, which is stored in the ZK-ROM. In our framework, the prover first computes all intermediate clauses locally. The two parties then interact according to the original protocol, comparing the final resolvent against the learned clause read from the ZK-ROM.

The lower bound of the number of possible proof structures in Eq.~\ref{eq:tree} is given by composing the sub-trees to create the structure of the full proof. Taking the example in Fig.~\ref{fig:tree}, we know that $R'=4$ and assume the proving sequence of sub-trees is blue, red, green, and purple. The verifier first attempts to determine the position of the green tree on the purple tree. Since $n_{\text{res}_4}=5$, and $j = 4 = R'$, there are 5 possible positions for the green tree to connect. Next, the verifier attempts to place the red tree. Because one position on the purple tree is occupied by the green tree, the red tree has $n_4+n_3=n_{\text{res}_4} + n_{\text{res}_3} - 1 = 5 + 2 - 1 = 6$ possible positions. Following this logic, the total number of potential proof structures, $N$, is calculated as:
\begin{equation*}
    N = 5*6*7 > (R'-1)! = 3!
\end{equation*}
This toy example demonstrates that the complexity of full proof structure reconstruction is lower bounded by the factorial of $(R'-1)$.

\section{An Example of the Conversion from a Verification Problem to Commitments of Polynomials}
\label{sec:app_example}
We first provide a simple example of the conversion from a verification problem to polynomials. We use the example in Fig.~\ref{fig:cnf}, which contains an AND gate and a property for verification. The formula is $\Phi=(a\lor \neg c)\land(b\lor \neg c)\land(\neg a \lor \neg b \lor c)\land c \land \neg a$, where $a$ and $c$ is public variable, while $b$ is secret.

We rename each variable with an index from $1$ to $3$, and let $l_i$ denote the literal corresponding to the $i$-th variable, such that $a=l_1,b=l_2$, and $c=l_3$. The formula is:
\begin{equation*}
    \Phi=(\neg l_1 \lor \neg l_2 \lor l_3)\land(l_1\lor \neg l_3)\land(l_2\lor \neg l_3)\land l_3 \land \neg l_1
\end{equation*}
The formula $\Phi$ is a conjunction of 5 clauses; therefore, we represent $\Phi$ with 5 polynomials. Take the first clause as an example; the encoded polynomial is:
\begin{equation*}
\begin{split}
    &\text{Poly}[(\neg l_1 \lor \neg l_2 \lor l_3)](x)\\
    =&(x+\phi(\neg l_1))(x+\phi(\neg l_2))(x+\phi(l_3))\\
    =&(x+(1+const_\text{lit})+const_\text{pub})(x+(H_k(2)+const_\text{lit}))\\
    &(x+3+const_\text{pub})
\end{split}
\end{equation*}

When we commit the polynomial, we transform the polynomial to its coefficient representation $a_3x^3 + a_2x^2 + a_1x^1 + a_0$ and commit it as $\{[a_3],[a_2],[a_1],[a_0]\}$. By committing all clauses in $\Phi$, we complete the commitment of the entire formula.

Next, an example of a resolution step encoded in polynomial relations is provided. Let the two antecedent clauses be $c_0=l_1\lor \neg l_3$ and $c_1=l_3$. Their encoded polynomials are:
\begin{equation*}
\begin{split}
    &\text{Poly}[c_0](x)=(x+1+const_\text{pub})(x+(3+const_\text{lit})+const_\text{pub})\\
    &\text{Poly}[c_1](x)=(x+3+const_\text{pub})
\end{split}
\end{equation*}

The witness clauses are $w_0=\emptyset$ and $w_1=l_1$, and the resolvent clause and the pivot are $c_r=l_1$ and $p=l_3$, respectively. Their corresponding polynomials are:
\begin{equation*}
\begin{split}
    &\text{Poly}[w_0](x)=1\\
    &\text{Poly}[w_1](x)=(x+1+const_\text{pub})\\
    &\text{Poly}[c_r](x)=(x+1+const_\text{pub})\\
    &\text{Poly}[p](x)=(x+3+const_\text{pub})\\
    &\text{Poly}[\neg p](x)=(x+(3+const_\text{lit})+const_\text{pub})
\end{split}
\end{equation*}

Following Eq.~\ref{eq:poly_relation}, we check a single resolution step by proving the following polynomial relation:
\begin{equation*}
\begin{split}
    &\text{Poly}[w_0](x)\cdot\text{Poly}[c_0](x)=\text{Poly}[c_r](x)\cdot\text{Poly}[\neg p](x)\\
    &\text{Poly}[w_1](x)\cdot\text{Poly}[c_1](x)=\text{Poly}[c_r](x)\cdot\text{Poly}[p](x)
\end{split}
\end{equation*}

\section{The data flow diagram of $\Pi_\text{ZK-CEC}$}

\begin{figure}[tbp]
\centerline{\includegraphics[width=5.6cm]{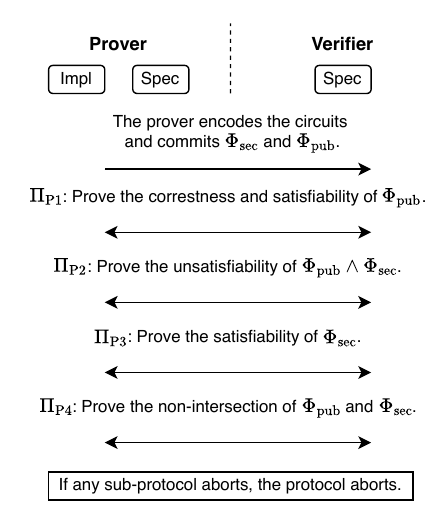}}
\caption{The data flow diagram of $\Pi_\text{ZK-CEC}$.}
\label{fig:flow}
\end{figure}

Fig.~\ref{fig:flow} demonstrates the data flow for our main protocol $\Pi_\text{ZK-CEC}$.

\section{Compliance with Open Science Policy}

Our research adheres to the Open Science policy to ensure transparency, reproducibility, and long-term impact. The full source code for our ZK-CEC protocol implementation is available at \url{https://github.com/NomadShen/ZKCEC}. All the circuit benchmarks and the scripts for proof generation can be accessed at \url{https://github.com/NomadShen/CNF-GEN}.





\end{document}